\documentclass[a4paper,11pt,leqno]{article}
\usepackage{graphicx}
\usepackage{subfigure}
\usepackage{amsmath}
\usepackage{amsfonts}
\usepackage{amssymb}

\usepackage{txfonts}
\usepackage{bm}
\usepackage{natbib}
\usepackage{multirow}
\usepackage{subfigure}
\usepackage{paralist}

\newcommand{\diff}{\mathrm{d}}
\newcommand{\reals}{\mathbb{R}}
\newcommand{\CC}{\mathbb{C}}

\newcommand{\E}{\operatorname{E}}

\DeclareMathOperator{\pr}{pr}
\renewcommand{\Pr}{\pr}

\DeclareMathOperator{\corr}{cor}
\DeclareMathOperator{\cov}{cov}

\newcommand{\dto}{\stackrel{d}{\rightarrow}}

\newlength{\LL}
\newlength{\LLL}
\newlength{\LLLL}

\begin{document}

%\jname{Biometrika}
%% The year, volume, and number are determined on publication
%\jyear{20??}
%\jvol{??}
%\jnum{?}
%% The \doi{...} and \accessdate commands are used by the production team
%\doi{10.1093/biomet/asm023}
%\accessdate{Advance Access publication on 30 June 2011}
%\copyrightinfo{\Copyright\ 2011 Biometrika Trust\goodbreak {\em Printed in Great Britain}}

%% These dates are usually set by the production team
%\received{January 2011}
%\revised{June 2011}

%% The left and right page headers are defined here:
%\markboth{Ingrid Hob\ae{}k Haff \and Johan Segers}{Empirical pair-copulae}

%% Here are the title, author names and addresses
\title{Nonparametric estimation of pair-copula constructions with the empirical pair-copula}

\author{I. HOB\AE{}K HAFF\footnote{Norwegian Computing Center, P.O. Box 114 Blindern, NO-0314 Oslo, Norway. E-mail: ingrid.haff@nr.no} 
\and J. SEGERS\footnote{ISBA, Universit\'e catholique de Louvain, Voie du Roman Pays 20, B-1348 Louvain-la-Neuve, Belgium. E-mail: johan.segers@uclouvain.be}}

\date{January 24, 2012}

\maketitle

\begin{abstract}
A pair-copula construction is a decomposition of a multivariate copula 
into a structured system, called regular vine, of bivariate copulae or 
pair-copulae. The standard practice is to model these pair-copulae 
parametrically, which comes at the cost of a large model risk, with errors 
propagating throughout the vine structure. The empirical pair-copula 
proposed in the paper provides a nonparametric alternative still achieving 
the parametric convergence rate. It can be used as a basis for inference on 
dependence measures, for selecting and pruning the vine structure, and for 
hypothesis tests concerning the form of the pair-copulae. \par \medskip
\end{abstract}

\section{Introduction}

Pair-copula constructions, introduced in \cite{joe:1996} and developed in 
\cite{bedford:cooke:2001, bedford:cooke:2002} and \cite{kurowicka:cooke:2006}, 
provide a flexible, but manageable way of modelling the dependence within a 
random vector. The crucial model assumption is that the copulae of certain 
bivariate conditional distributions do not depend on the value of the 
conditioning variable or vector. In this way, a copula in dimension $d$ is 
completely determined by the collection of pairwise connections between
conditional distributions for which the model assumption holds, called the 
vine structure of the copula, together with a set of $d(d-1)/2$ bivariate 
copulae, called pair-copulae. These are grouped into levels according to the 
number of conditioning variables of the corresponding conditional distributions, 
going from the ground level, comprising $d-1$ pair-copulae which are just
bivariate margins of the parent copula, up to the top level, consisting of the 
single copula being the copula of the remaining two variables, conditionally on 
the $d-2$ others.

Current practice is to model the pair-copulae parametrically, estimating the 
parameters with a composite or pseudo-likelihood method, that is either 
frequentistic, as in \cite{Vines} and \cite{haff:2012}, or Bayesian, as in 
\cite{min:czado:2010, min:czado:2011}. Fitting a pair-copula construction 
therefore requires the selection of $d(d-1)/2$ copula models. The recursive
dependence of inference concerning copulae at a certain level on the copulae 
fitted in the lower levels augments the model risk. Thus, bad model choices 
propagate errors throughout the vine structure.

In this paper, a nonparametric pair-copula estimator is proposed instead. Of 
course, if the parametric model is correctly specified, a parametric estimator 
will be more efficient. But the nonparametric method is more robust, as it 
does not rely on a parametric specification. The estimator is based on an 
idea similar to the empirical copula \citep{ruschendorf:1976,deheuvels:1979}, 
and is therefore called the empirical pair-copula. Although it joins conditional 
distributions, the empirical pair-copula still achieves the parametric rate, 
regardless of the number of conditioning variables, thanks to the model 
assumption that these copulae do not depend on the conditioning variable.

The empirical pair-copula yields nonparametric estimators of dependence 
measures such as conditional Spearman rank correlations. These estimates can 
safely be used in vine structure selection algorithms, yielding a nonparametric 
alternative to the procedure proposed in \cite{dissmann11}. Other applications 
of the empirical pair-copula concern testing for conditional independence at 
certain levels, aiming at pruning or truncating of the vine structure as in 
\cite{brechmann:czado:aas:2011}, as well as goodness-of-fit testing in 
combination with parametric methods. The new method is supported by extensive 
simulations and is illustrated by case studies involving financial and 
precipitation data.

\section{Pair-copula constructions}

First, let $F$ be the bivariate continuous distribution function of a 
random pair $(X_1, X_2)$, with margins $F_1$ and $F_2$ and copula $C$, 
that is,
\[
  F(x_1, x_2) = C \{ F_1(x_1), F_2(x_2) \}.
\]
The bivariate density $f$ of $F$ then satisfies
\[
  f(x_1, x_2) = c \{ F_1(x_1), F_2(x_2) \} \, f_1(x_1) \, f_2(x_2),
\]
where $f_1$ and $f_2$ denote the marginal density functions and $c$ is the 
copula density, and the conditional density of $X_1$, given $X_2 = x_2$, is
\begin{equation}
\label{eq:f12}
  f_{1|2}(x_1|x_2) = \frac{f(x_1, x_2)}{f_2(x_2)} = c(F_1(x_1), F_2(x_2)) \, f_1(x_1).
\end{equation}
The corresponding conditional distribution function satisfies
\begin{eqnarray}
\label{eq:F12}
  F_{1|2}(x_1|x_2) 
  &=& \int_{-\infty}^{x_1} f_{1|2}(z|x_2) \, \diff z 
  = \int_{-\infty}^{x_1} c \{ F_1(z), F_2(x_2) \} \, f_1(z) \, \diff z \nonumber \\
  &=& \int_0^{F_1(x_1)} c \{ u, F_2(x_2) \} \, \diff u 
  = \left. \frac{\partial}{\partial u_2} C(u_1, u_2) \right|_{(u_1, u_2) = (F_1(x_1), F_2(x_2))} \nonumber \\
  &=& C^{[2]} \{ F_1(x_1),  F_2(x_2)\}.
\end{eqnarray}

\begin{figure}
\begin{center}
\begin{tabular}{c@{\quad}c}
\includegraphics[width=0.42\textwidth]{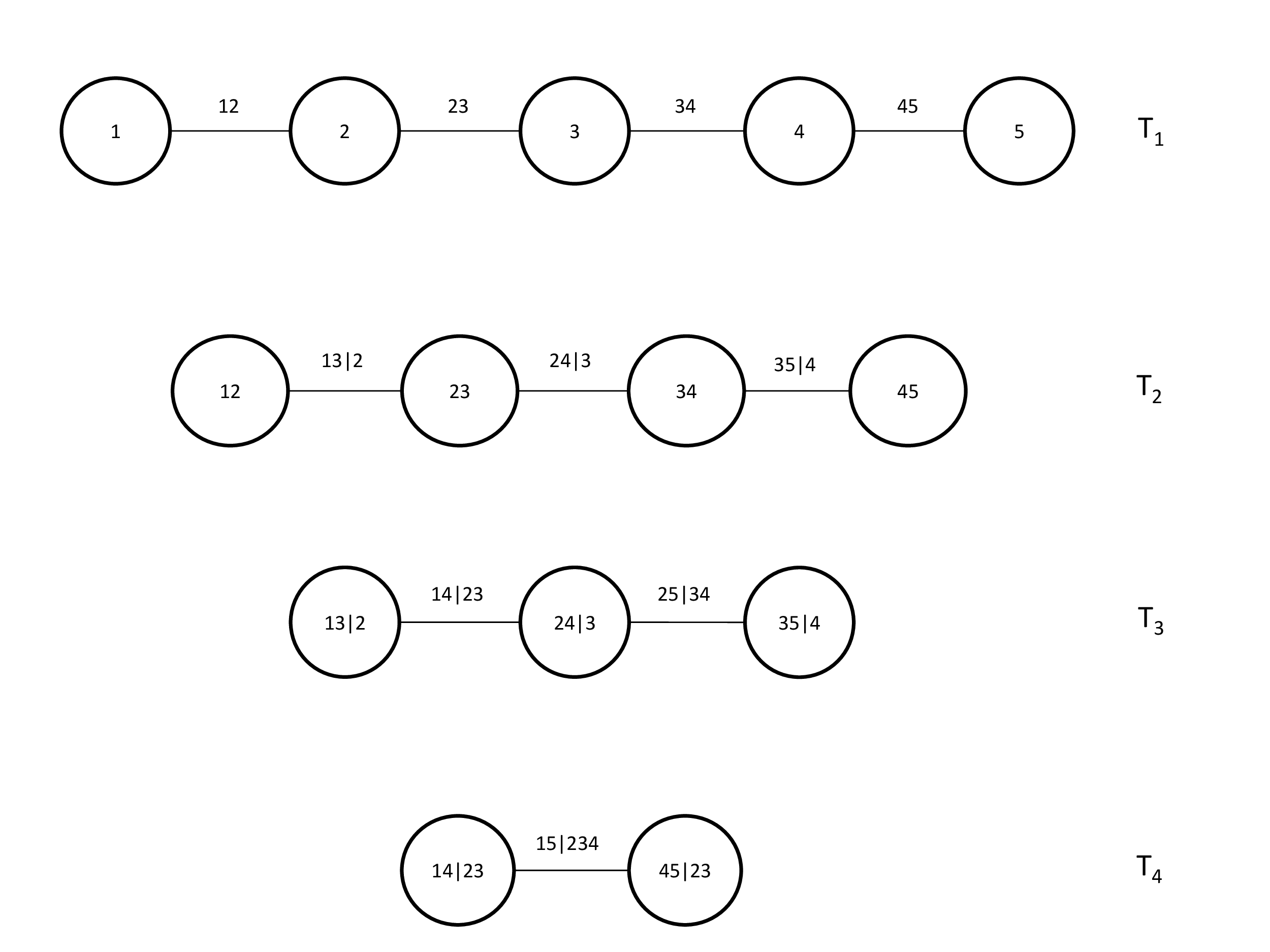}&
\includegraphics[width=0.42\textwidth]{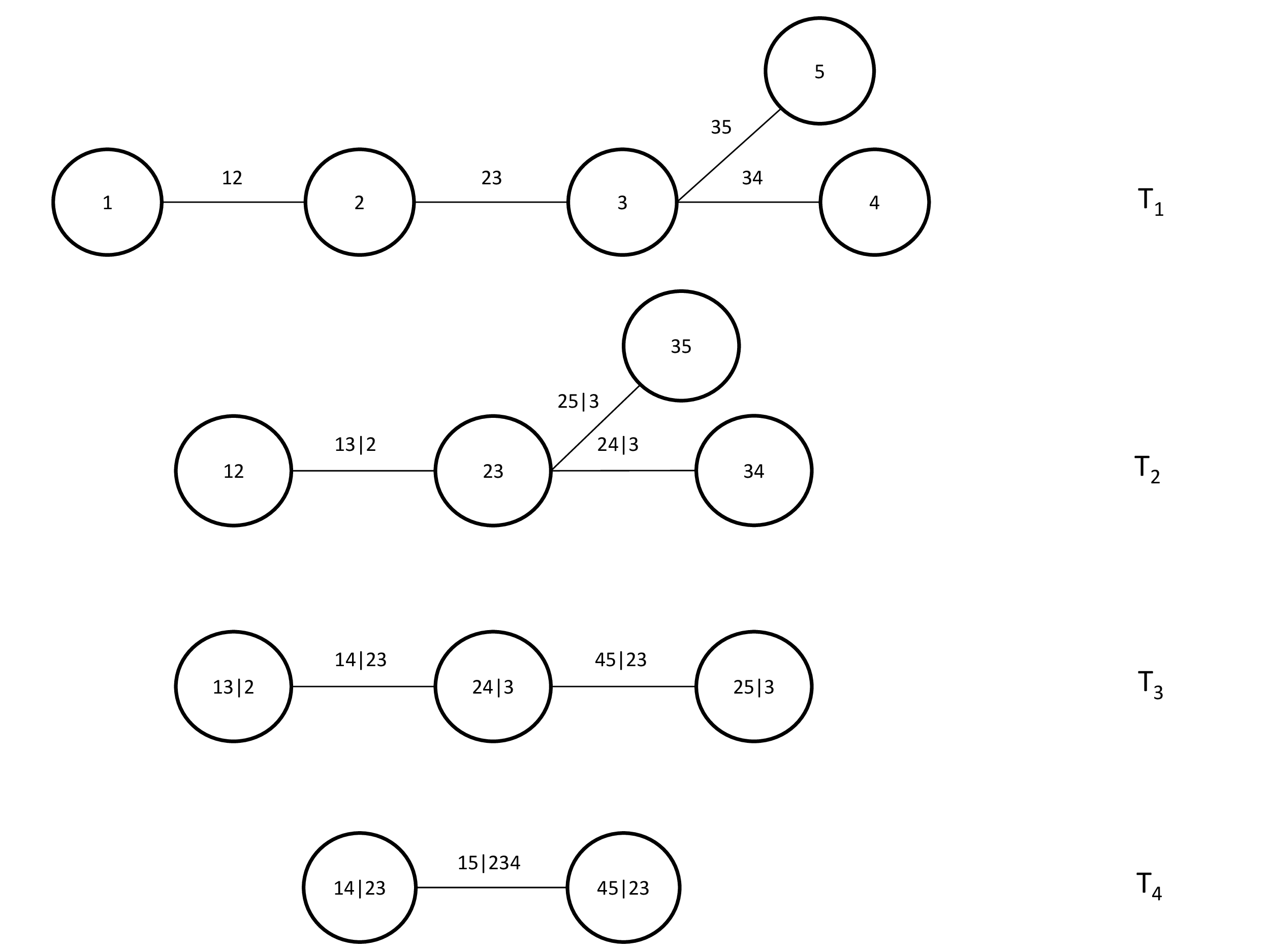}
\end{tabular}
\end{center}
\caption{\sl Examples of five-dimensional regular vines: a drawable vine (left) and a non-classified regular vine (right).}
\label{fig:DRvines}
\end{figure}

Next, let $f$ be the $d$-variate probability density function of the random 
vector $(X_1, \ldots, X_d)$ with $d \ge 3$. Let $i$ and $j$ be distinct 
elements of $\{1, \ldots, d\}$ and let $v$ be a non-empty subset of 
$\{1, \ldots, d\} \setminus \{i,j\}$. Write $X_v = (X_i : i \in v)$ and 
similarly for $x_v$. Applying \eqref{eq:f12} to the conditional density 
$f_{ij|v}(\cdot,\cdot|x_v)$ of the pair $(X_i, X_j)$, given $X_v = x_v$,
associated with the copula $C_{ij|v}(\cdot,\cdot|x_v)$ and its
density $c_{ij|v}(\cdot,\cdot|x_v)$,yields
\begin{equation}
\label{eq:fijv:general}
  f_{i|j \cup v}(x_i | x_j, x_v) = c_{ij|v} \{ F_{i|v}(x_i|x_v), F_{j|v}(x_j|x_v) | x_v\} \, f_{i|v}(x_i | x_v).
\end{equation}
From \eqref{eq:F12} it follows that
\begin{equation}
\label{eq:Fijv:general}
  F_{i|j \cup v}(x_i | x_j, x_v) = C_{ij|v}^{[2]} \{ F_{i|v}(x_i|x_v), F_{j|v}(x_j|x_v) | x_v \}.
\end{equation}
Equation \eqref{eq:fijv:general} provides a way to write $f_{i|j \cup v}$ in terms 
of $c_{ij|v}$ and $f_{i|v}$, with one variable less in the conditioning set. 
Applying this equation recursively to the terms on the right-hand side of the 
identity
\[
  f(x_1, \ldots, x_d) = f_1(x_1) \, f_{2|1}(x_2|x_1) \, \cdots \, f_{d|12\ldots(d-1)}(x_d|x_1, \ldots, x_{d-1})
\]
yields expressions of the form
\begin{equation}
\label{eq:PCC:general}
  f(x_1, \ldots, x_d) = \prod_{k=1}^d f_k(x_k) \, \prod_{\ell = 1}^{d-1} \prod_{(i,j,v)} c_{ij|v} \{ F_{i|v}(x_i|x_v), F_{j|v}(x_j|x_v) | x_v\}.
\end{equation}
The number of terms in the third product is equal to $d - \ell$. For each triple 
$(i,j,v)$ in the product, $v$ is a subset of $\{1, \ldots, d\} \setminus \{i, j\}$ 
with exactly $\ell - 1$ elements. The precise list of combinatorial rules that 
the system of triples $(i, j, v)$ must obey makes them constitute a regular vine as in
\cite{bedford:cooke:2001,bedford:cooke:2002}. Examples of two such structures 
in dimension five are given in Figure~\ref{fig:DRvines}.

Assume that for a specific choice of $(i, j, v)$, the copula density $c_{ij|v}$ 
does not depend on the value of the conditioning argument $x_v$, that is, 
$c_{ij|v}(u_i, u_j | x_v)$ is constant in $x_v$. Since the corresponding copula 
$C_{ij|v}(\cdot,\cdot|x_v)$ is equal to the joint distribution function of
%the random pair 
$(F_{i|v}(X_i|X_v), F_{j|v}(X_j|X_v))$ given $X_v = x_v$, we find 
that the random pair $(F_{i|v}(X_i|X_v), F_{j|v}(X_j|X_v))$ must be independent 
of the random vector $X_v$. Obviously, the converse must hold as well. In that 
case, equations \eqref{eq:fijv:general} and \eqref{eq:Fijv:general} simplify to
\begin{align}
\label{eq:fijv}
  f_{i|j \cup v}(x_i | x_j, x_v) &= c_{ij|v} \{ F_{i|v}(x_i|x_v), F_{j|v}(x_j|x_v) \} \, f_{i|v}(x_i | x_v), \\
\label{eq:Fijv}
  F_{i|j \cup v}(x_i | x_j, x_v) &= C_{ij|v}^{[2]} \{ F_{i|v}(x_i|x_v), F_{j|v}(x_j|x_v) \}.
\end{align}
If it is true for all triples $(i, j, v)$ in the regular vine in 
\eqref{eq:PCC:general}, we arrive at the pair-copula construction
\citep{joe:1996,kurowicka:cooke:2006}
\begin{equation}
\label{eq:PCC}
  f(x_1, \ldots, x_d) = \prod_{k=1}^d f_k(x_k) \, \prod_{\ell = 1}^{d-1} \prod_{(i,j,v)} c_{ij|v} \{ F_{i|v}(x_i|x_v), F_{j|v}(x_j|x_v) \},
\end{equation}
that provides a decomposition of a $d$-variate density in terms of $d$ 
univariate and $d(d-1)/2$ bivariate copula densities. The pair-copula 
construction corresponding to the drawable vine in the left panel of 
Figure~\ref{fig:DRvines} is
\begin{eqnarray*}
\lefteqn{
c \{ F_{1}(x_{1}),\ldots,F_{5}(x_{5}) \}
} \\
&=& c_{12} \{ F_{1}(x_{1}),F_{2}(x_{2}) \} \ c_{23} \{ F_{2}(x_{2}),F_{3}(x_{3}) \} \ c_{34} \{ F_{3}(x_{3}),F_{4}(x_{4}) \} \ c_{45}\{F_{4}(x_{4}),F_{5}(x_{5})\}\\
&& c_{13|2} \{ F_{1|2}(x_{1}|x_{2}),F_{3|2}(x_{3}|x_{2}) \} \ c_{24|3} \{ F_{2|3}(x_{2}|x_{3}),F_{4|3}(x_{4}|x_{3}) \} \ c_{35|4} \{ F_{3|4}(x_{3}|x_{4}),F_{5|4}(x_{5}|x_{4}) \}\\
&& c_{14|23} \{ F_{1|23}(x_{1}|x_{2},x_{3}),F_{4|23}(x_{4}|x_{2},x_{3}) \} \ c_{25|34} \{ F_{2|34}(x_{2}|x_{3},x_{4}),F_{5|34}(x_{5}|x_{3},x_{4}) \} \\
&& c_{15|234} \{ F_{1|234}(x_{1}|x_{2},x_{3},x_{4}),F_{5|234}(x_{5}|x_{2},x_{3},x_{4}) \}.
\end{eqnarray*}

The assumption that the pair-copulae do not depend on the value of the 
conditioning argument is a nonparametric shape constraint, that is satisfied 
for instance by the multivariate Student's t and Clayton copulae \citep{PCCcover}. 
Even if the assumption does not hold in general, it still provides a reasonable 
approximation to the true distribution in many cases.

\section{Empirical pair-copula}

\subsection{Estimator}

Let $X_t = (X_{1t}, \ldots, X_{dt})$, for $t = 1, \ldots, n$, be a $d$-variate random 
sample from a distribution function $F$ with density $f$, admitting a pair-copula 
construction \eqref{eq:PCC} with a known regular vine structure. Choice of the vine 
structure is a difficult problem, which we will address in Section~\ref{subsec:vine}. 
Consider the ground level normalized ranks
\[
  \hat{U}_{itn} = \frac{1}{n+1} \sum_{s=1}^n I(X_{is} \le X_{it}) \qquad (i = 1, \ldots, d;\ t = 1, \ldots, n).
\]
The ground level empirical pair-copula is simply the classical empirical copula
\[
  \hat{C}_{ij,n}(u_i, u_j) = \frac{1}{n} \sum_{t=1}^n I \bigl( \hat{U}_{itn} \le u_i, \, \hat{U}_{jtn} \le u_j \bigr) \qquad (i, j \in \{1, \ldots, d\};\ i \ne j).
\]
Use finite differencing to obtain an estimator of the conditional distribution function: 
writing $\hat{C}_{ij,n}(A) = n^{-1} \sum_{t=1}^n I \{ ( \hat{U}_{itn}, \hat{U}_{jtn} ) \in A \}$ 
for $A \subset \reals^{2}$ and given a bandwidth $h > 0$, first put
\begin{eqnarray}
\label{eq:C2}
  \hat{C}_{ij,n}^{[2]} (u_i, u_j)
  &=& \frac{\hat{C}_{ij,n} \bigl( [0, u_i] \times [u_j - h, u_j + h] \bigr)}{\hat{C}_{ij,n} \bigl( [0, 1] \times [u_j - h, u_j + h] \bigr)} \\
%   &= \frac{\hat{C}_{ij,n}(u_i, u_j + h) - \hat{C}_{ij,n}(u_i, (u_j - h)-)}{\hat{C}_{ij,n}(1, u_j + h) - \hat{C}_{ij,n}(1, (u_j - h)-)} \\
  &=& \frac{\sum_{s=1}^n I \bigl( \hat{U}_{isn} \le u_i, \, |\hat{U}_{jsn} - u_j| \le h \bigr)}{\sum_{s=1}^n I \bigl( |\hat{U}_{jsn} - u_j| \le h \bigr)}, \nonumber
\end{eqnarray}
and then, following \eqref{eq:F12},
\begin{eqnarray}
\label{eq:condground}
  \hat{F}_{i|j,n}(X_{it}|X_{jt}) 
  &=& \hat{C}_{ij,n}^{[2]}(\hat{U}_{itn}, \hat{U}_{jtn}) \\
  &=& \frac{\sum_{s=1}^n I \bigl( \hat{U}_{isn} \le \hat{U}_{itn}, \, |\hat{U}_{jsn} - \hat{U}_{jtn}| \le h \bigr)}%
  {\sum_{s=1}^n I \bigl( |\hat{U}_{jsn} - \hat{U}_{jtn}| \le h \bigr)}.  \nonumber
\end{eqnarray}
The denominator of \eqref{eq:condground} is approximately equal to $2nh$, except at the borders, 
where it is smaller, providing a boundary correction. As the smoothing step in \eqref{eq:C2} takes 
place on a uniform $(0, 1)$ scale, the choice of bandwidth $h$ does not depend on the marginal 
distributions; in fact, \eqref{eq:condground} is a kind of nearest-neighbour estimator. Bandwidth 
selection will be addressed in Section~\ref{subsec:proofbysimul}, where it will be seen that a 
slight degree of undersmoothing is advizable.

For higher levels, we proceed recursively, exploiting the assumption that the pair-copulae do 
not depend on the value of the conditioning argument. The unwinding of the recursion depends on 
the given vine structure. Let $(i, j, v)$ be a triple in the vine decomposition 
\eqref{eq:PCC}; in particular, $v$ is a subset of $\{1, \ldots, d\}$ and $i$ and $j$ are 
distinct elements of $\{1, \ldots, d\} \setminus v$. Suppose that the estimators 
$\hat{F}_{k|v,n}(X_{kr}| X_{vr})$ have been defined for $k \in \{i, j\}$ and $r = 1, \ldots, n$; 
here $X_{vr}$ denotes the random vector $(X_{mr}: m \in v)$. The normalized ranks of the estimated 
conditional probabilities are
\begin{multline}
\label{eq:hatUn}
  \hat{U}_{k|v,tn} = \frac{1}{n+1} \sum_{r=1}^n I \{ \hat{F}_{k|v,n}(X_{kr}| X_{vr}) \le \hat{F}_{k|v,n}(X_{kt}|X_{vt}) \} \\ (k \in \{i,j\};\ t = 1, \ldots, n).
\end{multline}
The empirical pair-copula is then defined by
\begin{equation}
\label{eq:emppaircop}
  \hat{C}_{ij|v,n}(u_i, u_j) = \frac{1}{n} \sum_{s=1}^n I \bigl( \hat{U}_{i|v, sn} \le u_i, \, \hat{U}_{j|v, sn} \le u_j \bigr).
\end{equation}
Again, apply finite differencing to get hold on the conditional distributions: first,
\begin{eqnarray}
\label{eq:emppaircopdiff}
  \hat{C}_{ij|v,n}^{[2]}(u_i, u_j)
  &=& \frac{\hat{C}_{ij|v,n} \bigl( [0, u_i] \times [u_j - h, u_j + h] \bigr)}{\hat{C}_{ij|v,n} \bigl( [0, 1] \times [u_j - h, u_j + h] \bigr)}\\
  &=& \frac{\sum_{s=1}^n I \bigl( \hat{U}_{i|v,sn} \le u_i, \, |\hat{U}_{j|v,sn} - u_j| \le h \bigr)}{\sum_{s=1}^n I \bigl( |\hat{U}_{j|v,sn} - u_j| \le h \bigr)}\nonumber
\end{eqnarray}
and then, following \eqref{eq:Fijv},
\begin{eqnarray}
\label{eq:condhigh}
  \hat{F}_{i|v \cup j,n}(X_{it}|X_{v \cup j, t}) 
  &=& \hat{C}_{ij,n}^{[2]}(\hat{U}_{i|v,tn}, \hat{U}_{j|v,tn})  \nonumber\\
  &=& \frac{\sum_{s=1}^n I \bigl( \hat{U}_{i|v,sn} \le \hat{U}_{i|v,tn}, \, |\hat{U}_{j|v,sn} - \hat{U}_{j|v,tn}| \le h \bigr)}%
  {\sum_{s=1}^n I \bigl( |\hat{U}_{j|v,sn} - \hat{U}_{j|v,tn}| \le h \bigr)}.
\end{eqnarray}
We proceed this way, recursively from the ground level, $\ell = 1$, where $v$ is the 
empty set, to the top level, $\ell = d-1$, with $v$ consisting of $d-2$ elements, adding 
a variable for each level.

The empirical pair-copula estimates the pair-copula distribution functions. 
\cite{kolbjstien08} propose a nonparametric estimator for the pair-copula density, 
with variables transformed to the Gaussian rather than the uniform domain, to mitigate 
boundary effects.

\subsection{Asymptotic distribution}

Let $(i,j,v)$ be a triple in the vine decomposition \eqref{eq:PCC}. Because of the 
assumption that the copula of the conditional distribution of $(X_i, X_j)$ given 
$X_v = x_v$ does not depend on the value of $x_v$, the pair-copula $C_{ij|v}$ is in 
fact equal to the unconditional distribution function of the random pair 
$(F_{i|v}(X_i|X_v), F_{j|v}(X_j|X_v))$:
\begin{eqnarray*}
  \lefteqn{
  \Pr \{ F_{i|v}(X_i|X_v) \le u_i, \, F_{j|v}(X_j|X_v) \le u_j \}
  } \\
  &=& \int \Pr \{ F_{i|v}(X_i|X_v) \le u_i, \, F_{j|v}(X_j|X_v) \le u_j \mid X_v = x_v \} \, f_v(x_v) \, \diff x_v \\
  &=& \int C_{ij|v}(u_i, u_j|x_v) \, f_v(x_v) \, \diff x_v \\
  &=& \int C_{ij|v}(u_i, u_j) \, f_v(x_v) \, \diff x_v
  = C_{ij|v}(u_i, u_j).
\end{eqnarray*}
Therefore, it is reasonable to expect that it can be estimated at the parametric rate 
$O_p(n^{-1/2})$.

Define the random variables 
\begin{equation}
\label{eq:U}
  U_{k|v,t} = F_{k|v}(X_{kt} | X_{vt}) \qquad (k \in \{i,j\};\ t = 1, \ldots, n).
\end{equation}
We conjecture that under suitable smoothness conditions on the copula density $c$ and 
growth conditions on the bandwidth sequence $h = h_n$, the empirical pair-copula 
\eqref{eq:emppaircop} satisfies
\begin{eqnarray}
\label{eq:emppaircopproc}
  \CC_{ij|v,n}(u_i, u_j) 
  &=& n^{1/2} \{ \hat{C}_{ij|v,n}(u_i, u_j) - C_{ij|v}(u_i, u_j) \} \nonumber \\
  &=& n^{-1/2} \sum_{t=1}^n \{ I ( U_{i|v,t} \le u_i, U_{j|v,t} \le u_j ) - C_{ij|v}(u_i, u_j) \} \nonumber \\
  && \quad \mbox{} - C_{ij|v}^{[1]}(u_i, u_j) \, n^{-1/2} \sum_{t=1}^n \{ I ( U_{i|v,t} \le u_i ) - u_i \} \nonumber \\
  && \quad \mbox{} - C_{ij|v}^{[2]}(u_i, u_j) \, n^{-1/2} \sum_{t=1}^n \{ I ( U_{j|v,t} \le u_j ) - u_j \} + o_p(1).
\end{eqnarray}
The expansion \eqref{eq:emppaircopproc} is suggested by tedious calculations and  
supported by extensive simulations summarized in Section~\ref{subsec:proofbysimul}.

Incidentally, the right-hand side of \eqref{eq:emppaircopproc} coincides with the expansion 
for the empirical copula process of the unobservable random pairs $(U_{i|v,t}, U_{j|v,t})$, 
for $t = 1, \ldots, n$. This empirical copula would arise if the estimated conditional 
distribution functions $\hat{F}_{k|v,n}$ in equation~\eqref{eq:hatUn} were replaced by the true 
ones, $F_{k|v}$, with normalized ranks
\begin{multline}
\label{eq:hatU}
  U_{k|v,tn} = \frac{1}{n+1} \sum_{r=1}^n I \{ F_{k|v,n}(X_{kr}| X_{vr}) \le F_{k|v,n}(X_{kt}|X_{vt}) \} \\ (k \in \{i,j\};\ t = 1, \ldots, n),
\end{multline}
and the empirical copula
\begin{equation}
\label{eq:empcop}
  C_{ij|v,n}(u_i, u_j) = \frac{1}{n} \sum_{t=1}^n I \bigl( U_{i|v,tn} \le u_i, \, U_{j|v,tn} \le u_j \bigr),
\end{equation}
without hats. By theory going back to \cite{ruschendorf:1976} and \cite{stute:1984}, 
equation~\eqref{eq:emppaircopproc} holds when the empirical pair-copula $\hat{C}_{ij|v,n}$ 
\eqref{eq:emppaircop} is replaced by the empirical copula $C_{ij|v,n}$ \eqref{eq:empcop}. As 
we are working with the ranks of the variables $\hat{F}_{k|v,n}(X_{kt}|X_{vt})$ ($t = 1, \ldots, n$), 
rather than the values themselves, it is intuitively not unreasonable to expect that replacing 
$\hat{F}_{k|v,n}$ by $F_{k|v,n}$ makes no difference asymptotically. For some recent references 
on the empirical copula see \cite{fermanian:radulovic:wegkamp:2004}, \cite{tsukahara:2005}, 
\cite{vandervaart:wellner:2007}, and \cite{segers:2012}. 

The expansion in \eqref{eq:emppaircopproc} implies that the empirical pair-copula is asymptotically 
normal,
\begin{equation}
\label{eq:asnorm}
  n^{1/2} \{ \hat{C}_{ij|v,n}(u_i, u_j) - C_{ij|v}(u_i, u_j) \} \dto N \bigl(0, \sigma_{ij|v}^2(u_i, u_j) \bigr) \qquad (n \to \infty),
\end{equation}
with asymptotic variance equal to
\begin{multline}
\label{eq:asvar}
  \sigma_{ij|v}^2(u_i, u_j) 
%   &= \var \bigl\{ I ( U_{i|v,t} \le u_i, \, U_{j|v,t} \le u_j )
%   - C_{ij|v}^{[1]}(u_i, u_j) \, I (U_{i|v,t} \le u_i) 
%   - C_{ij|v}^{[2]}(u_i, u_j) \, I (U_{j|v,t} \le u_j) \bigr\} \\
  = C_{ij|v} (1 - C_{ij|v}) + (C_{ij|v}^{[1]})^2 \, u_i(1-u_i) + (C_{ij|v}^{[2]})^2 \, u_j(1-u_j) \\
  - C_{ij|v}^{[1]} \, C_{ij|v}(1 - u_i) - C_{ij|v}^{[2]} \, C_{ij|v}(1 - u_j) + C_{ij|v}^{[1]} \, C_{ij|v}^{[2]} \, (C_{ij|v} - u_i u_j),
\end{multline}
where the arguments $(u_i, u_j)$ of $C_{ij|v}$ and its partial derivatives have been suppressed 
in the notation. Replacing $C_{ij|v}$ and its derivatives by the estimators 
\eqref{eq:emppaircop}--\eqref{eq:emppaircopdiff} yields a plug-in estimator 
$\hat{\sigma}_{ij|v}^2(u_i, u_j)$ for the asymptotic variance.

\subsection{Resampling}
\label{s:resampling}

The empirical pair-copula will most naturally be used for interval estimation and hypothesis 
tests. To be able to derive critical values, one needs resampling procedures. Here we propose 
the multiplier bootstrap for the empirical pair-copula process in \eqref{eq:emppaircopproc}. 
It resembles the approach for the ordinary empirical copula process, proposed by 
\cite{remillard:scaillet:2009} and studied in \cite{bucher:dette:2010} and \cite{segers:2012}.

Consider first the bivariate empirical process
\[
  \alpha_{ij|v,n}(u_i, u_j) = n^{-1/2} \sum_{t=1}^n \bigl\{ I ( U_{i|v,t} \le u_i, \, U_{j|v,t} \le u_j ) - C_{ij|v}(u_i, u_j) \bigr\},
\]
based upon the random variables $U_{k|v,t}$ from \eqref{eq:U}. Let $\xi_1, \ldots, \xi_n$ be 
independent and identically distributed random variables, independent of the original sample 
$X_1, \ldots, X_n$, with mean zero, unit variance, and a finite absolute moment of some order 
larger than two, for instance from the standard normal distribution. By Lemma~A.1 in 
\cite{remillard:scaillet:2009}, the process
\begin{eqnarray*}
  \alpha_{ij|v,n}'(u_i, u_j) 
  &=& n^{-1/2} \sum_{t=1}^n \xi_t \, \bigl\{ I ( U_{i|v,tn} \le u_i, \, U_{j|v,tn} \le u_j ) - C_{ij|v,n}(u_i, u_j) \bigr\} \\
  &=& n^{-1/2} \sum_{t=1}^n \bigl( \xi_t - \bar{\xi}_n \bigr) \, I ( U_{i|v,tn} \le u_i, \, U_{j|v,tn} \le u_j )
\end{eqnarray*}
is an asymptotically independent distributional copy of $\alpha_{ij|v,n}$. We therefore propose
\[
  \hat{\alpha}_{ij|v,n}'(u_i, u_j)
  = n^{-1/2} \sum_{t=1}^n \bigl( \xi_t - \bar{\xi}_n \bigr) \, I \bigl( \hat{U}_{i|v,tn} \le u_i, \, \hat{U}_{j|v,tn} \le u_j \bigr)
\]
as a bootstrap resample of $\alpha_{ij|v,n}(u_i, u_j)$. In view of equation~\eqref{eq:emppaircopproc}, 
we then suggest resampling $\CC_{ij|v,n}(u_i, u_j)$ by
\begin{equation}
\label{eq:empcopprocmult}
  \hat{\alpha}_{ij|v,n}'(u_i, u_j) \\
  - \hat{C}_{ij|v,n}^{[1]} (u_i, u_j) \, \hat{\alpha}_{ij|v,n}'(u_i, 1)
  - \hat{C}_{ij|v,n}^{[2]} (u_i, u_j) \, \hat{\alpha}_{ij|v,n}'(1, u_j).
\end{equation}

Repeating the procedure for $B$ independent rows $\xi_{b1}, \ldots, \xi_{bn}$, with 
$b \in \{1,\ldots,B\}$, gives $B$ approximately independent distributional copies of 
$\alpha_{ij|v,n}(u_i, u_j)$, and thus of $\CC_{ij|v,n}(u_i, u_j)$. The pointwise sample 
variance of these $B$ resamples may serve as an alternative estimator 
$\tilde{\sigma}_{ij|v}^{2}(u_i, u_j)$ of~\eqref{eq:asvar}. Two-sided asymptotic confidence 
intervals for $C_{ij|v}(u_i, u_j)$ with confidence level $1 - \alpha$ can be obtained by
\begin{equation}
\label{eq:cimultboot}
  \bigl[ \hat{C}_{ij|v,n} - n^{-1/2} \hat{q}_{n,1-\alpha/2}, \, \hat{C}_{ij|v,n} - n^{-1/2} \hat{q}_{n,\alpha/2} \bigr],
\end{equation}
where $\hat{q}_{n,\beta}$ is either the bootstrap estimate of the $\beta$-quantile of 
$\CC_{ij|v,n}(u_i, u_j)$, that is, the $\beta$-percentile of the bootstrap samples, or 
$\tilde{\sigma}_{ij|v} \, \Phi^{-1}(\beta)$, where $\Phi^{-1}$ is the quantile function 
of the standard normal distribution.

\subsection{Simulation studies of the asymptotic distribution}
\label{subsec:proofbysimul}

We have substantiated the conjectured expansion \eqref{eq:emppaircopproc}, limiting 
distribution \eqref{eq:asnorm}, with \eqref{eq:asvar}, and the resampling procedure 
from Section \ref{s:resampling} through simulation. The study includes different 
types of structures, pair-copula models and parameter values. In each experiment, we 
have generated $1,000$ samples of size $n$ from the model in question, with $n$ 
ranging from $100$ to $100,000$. For each sample, we have computed 
$\CC_{ij|v,n}(u_i, u_j)$, as well as the absolute difference between $\CC_{ij|v,n}(u_i, u_j)$
and the expansion \eqref{eq:emppaircopproc}, in a set of chosen points $(u_{i},u_{j})$, 
at given levels of the structure. The models of the study are five-dimensional and
comprise the drawable and regular vines of Figure \ref{fig:DRvines}, and a canonical
vine, which is another special case. The latter two are Gaussian, and the former 
either Gaussian, Student's t or Gumbel. The parameters of all the Gumbel copulae are 
$\theta=1$$\cdot$$5$. Further, the Gaussian and Student's t correlations are  
$\rho$, $\rho/(1+\rho)$, $\rho/(1+2\rho)$ and $\rho/(1+3\rho)$ at the first, second, 
third and fourth level, respectively, with $\rho=0$$\cdot$$2$, $0$$\cdot$$5$, 
$0$$\cdot$$8$. The corresponding degrees of freedom of the latter are $\nu$, $\nu+1$, 
$\nu+2$ and $\nu+3$, with $\nu=6$. 

\begin{figure}
\begin{center}
\begin{tabular}{c@{\quad}c@{\quad}c}
\vspace{-5mm}
\rotatebox{270}{\includegraphics[width=0.2\linewidth]{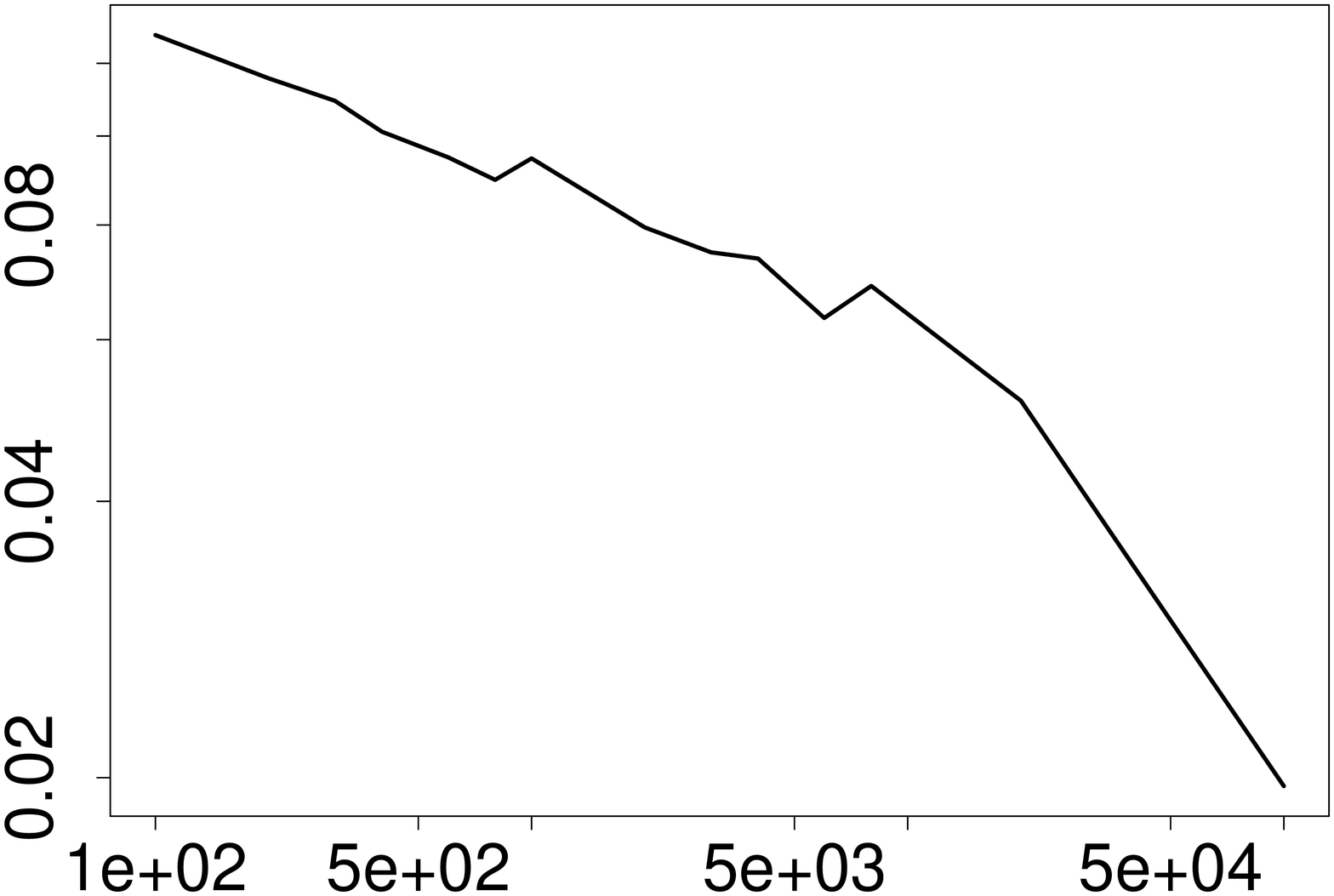}}&
\rotatebox{270}{\includegraphics[width=0.2\linewidth]{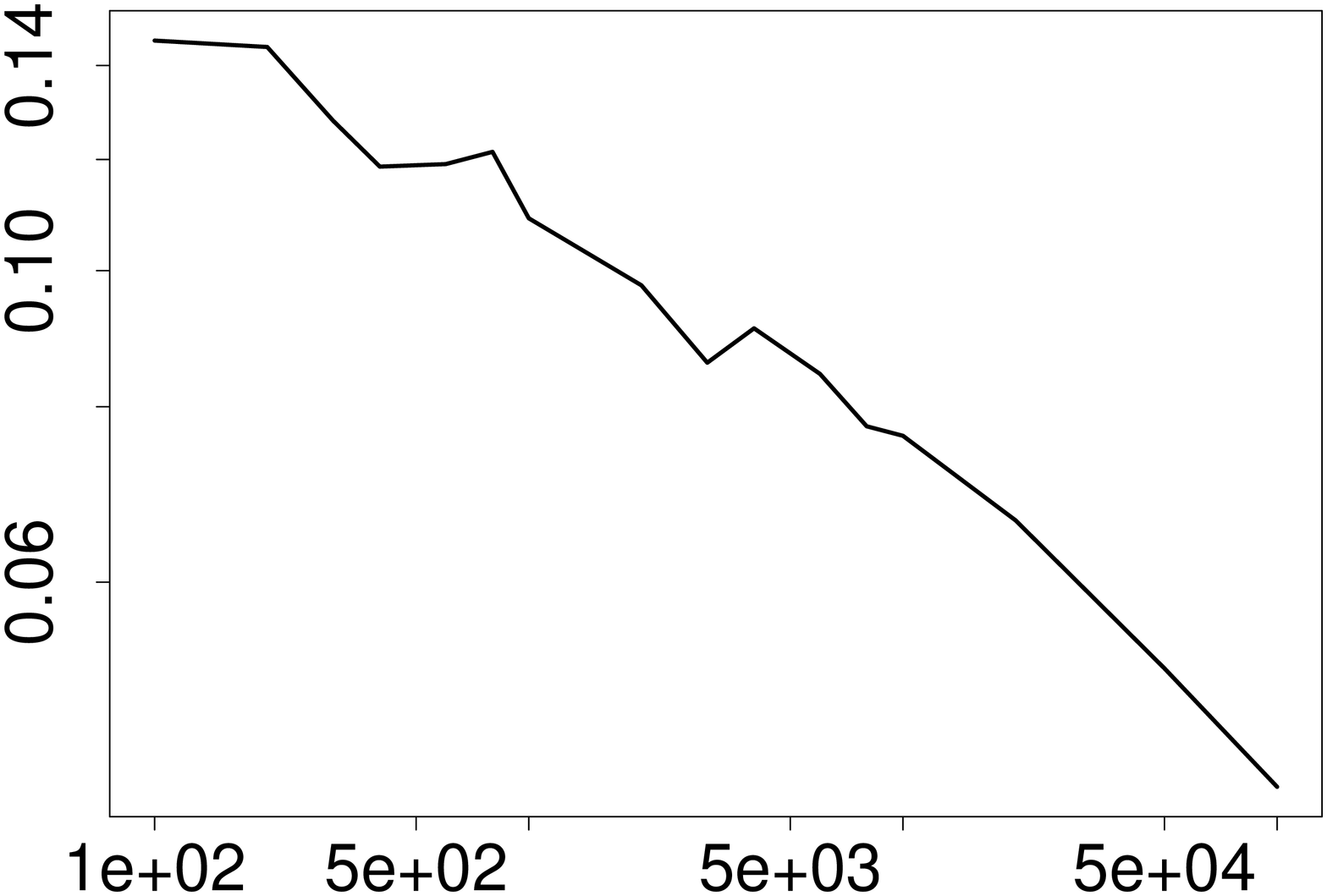}}&
\rotatebox{270}{\includegraphics[width=0.2\linewidth]{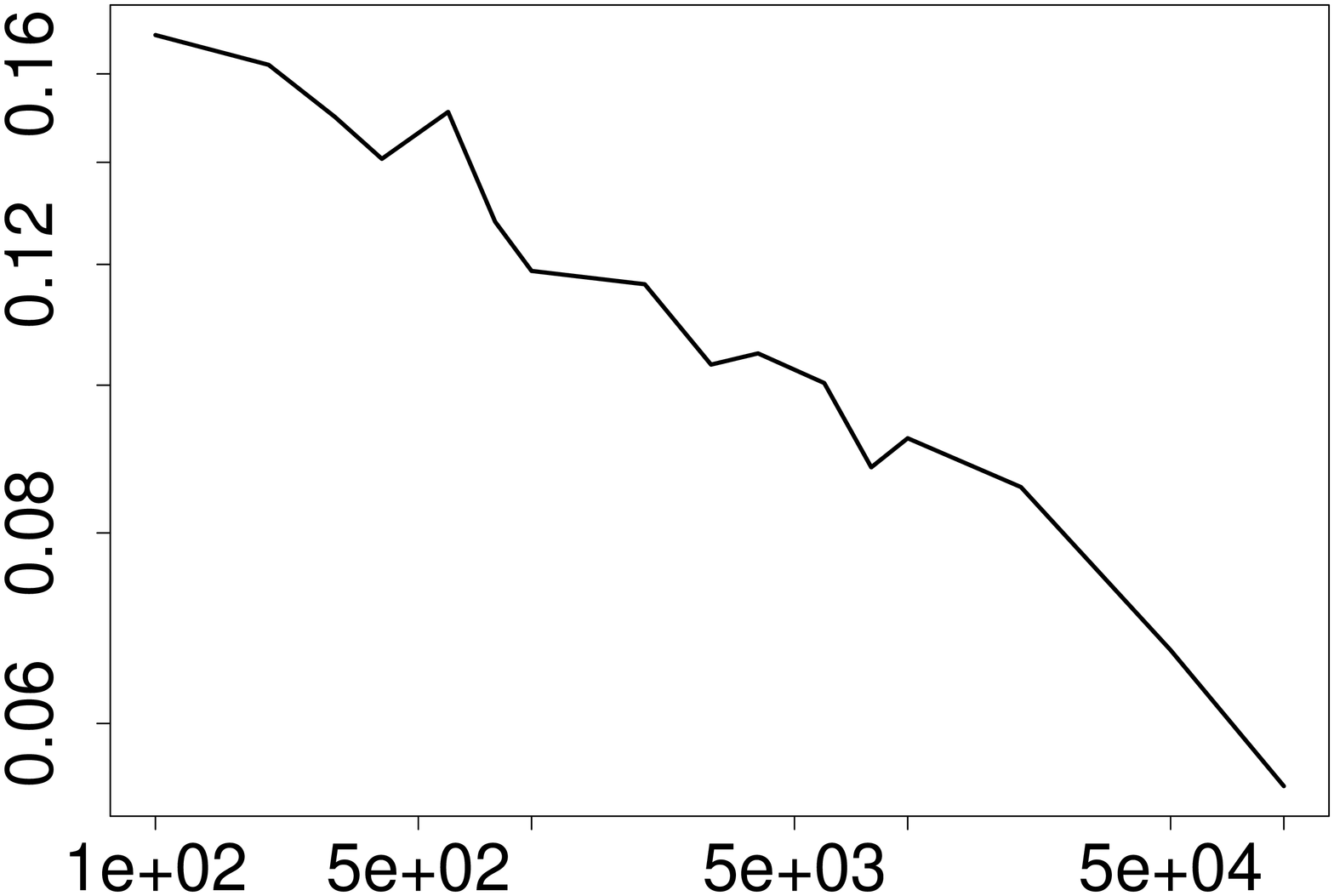}}\\
\vspace{-5mm}
\rotatebox{270}{\includegraphics[width=0.2\linewidth]{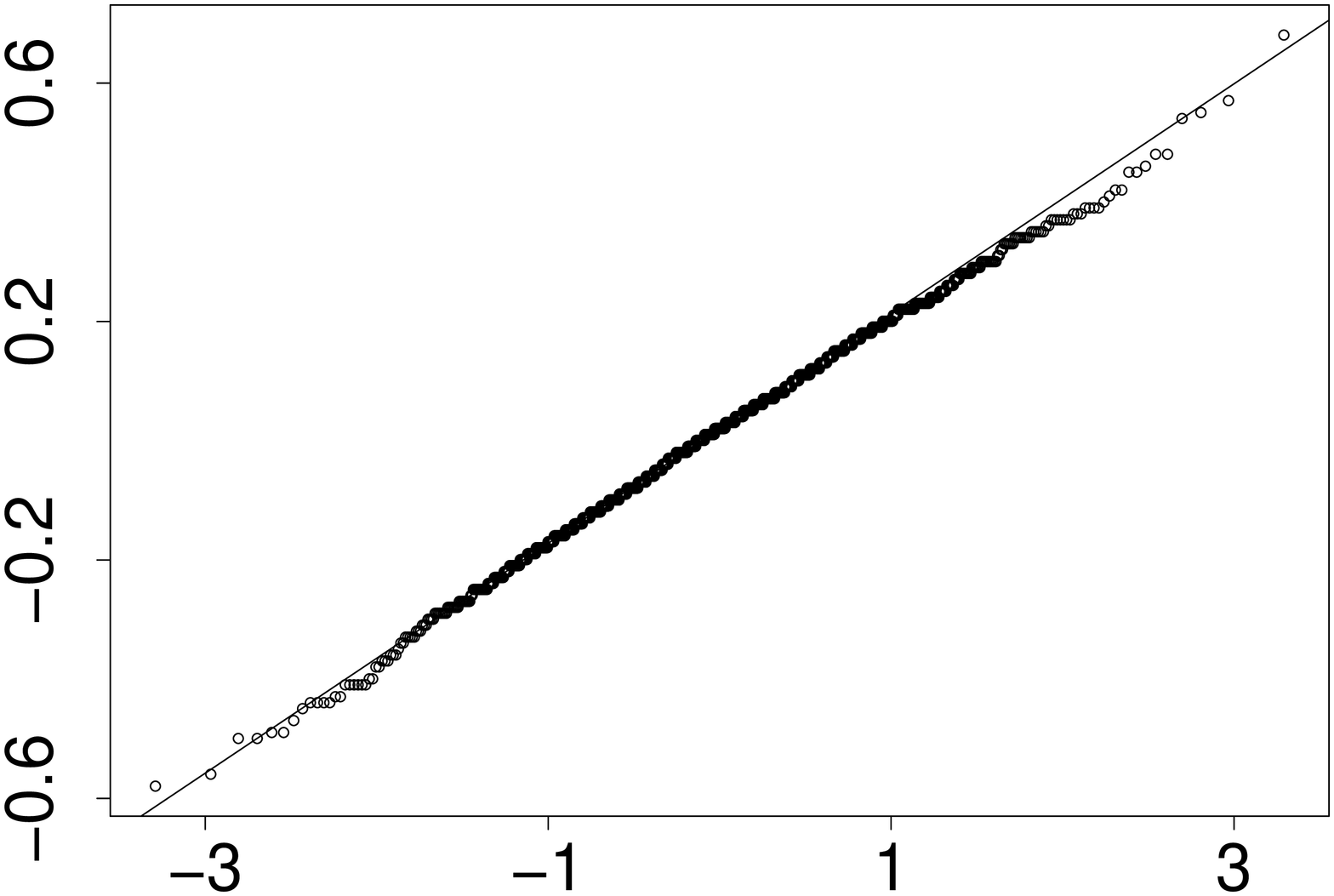}}&
\rotatebox{270}{\includegraphics[width=0.2\linewidth]{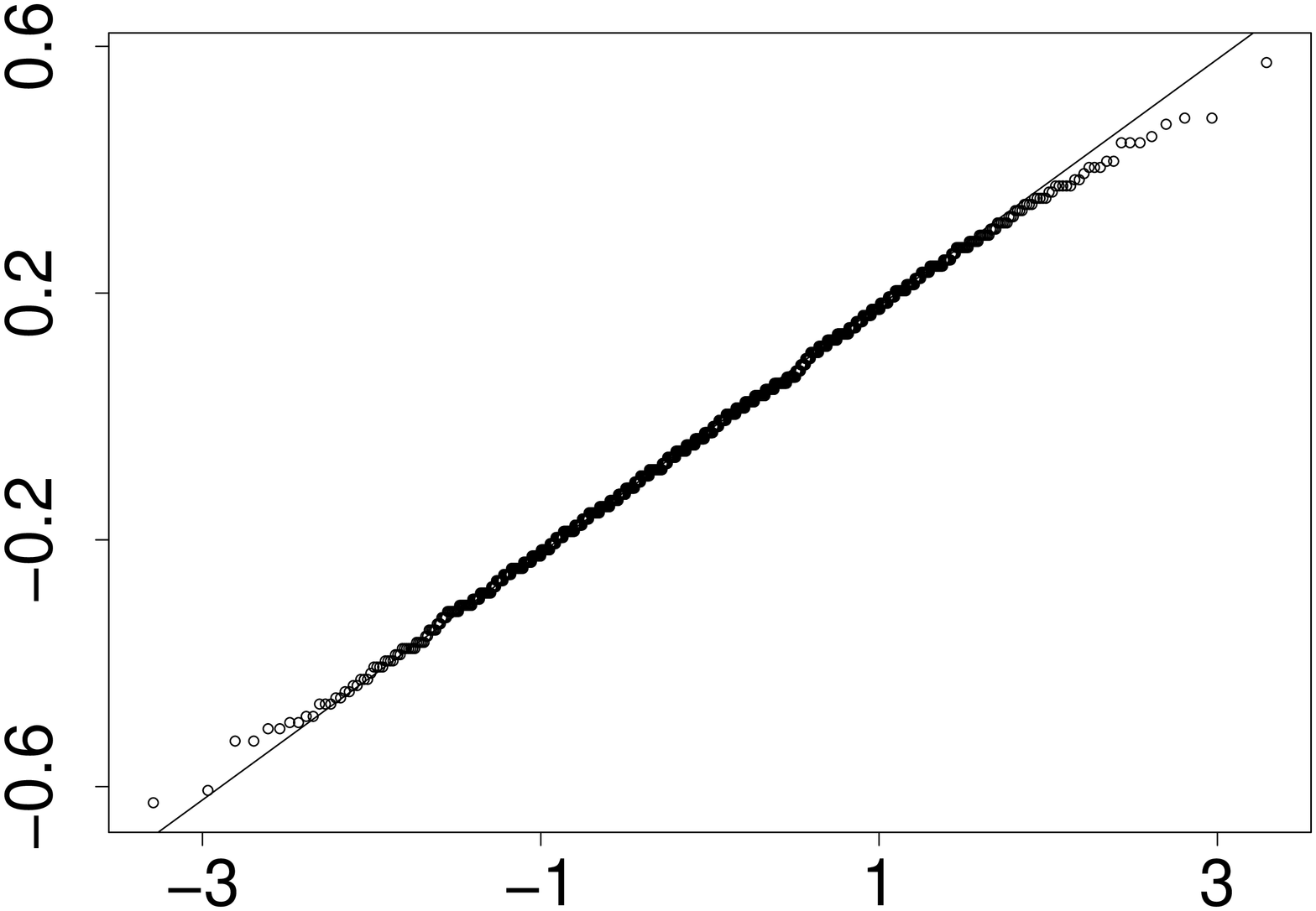}}&
\rotatebox{270}{\includegraphics[width=0.2\linewidth]{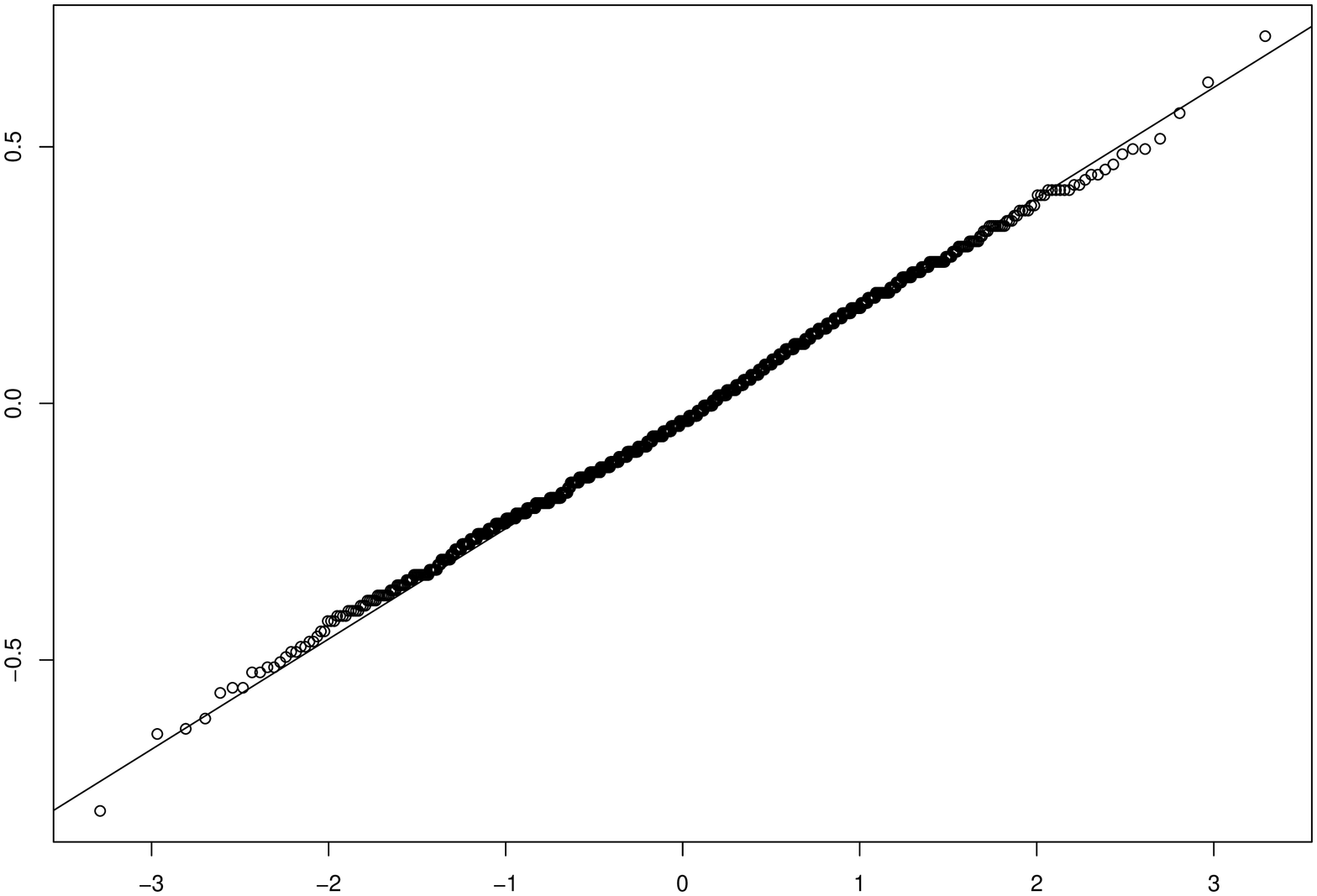}}\\
\vspace{-5mm}
\rotatebox{270}{\includegraphics[width=0.2\linewidth]{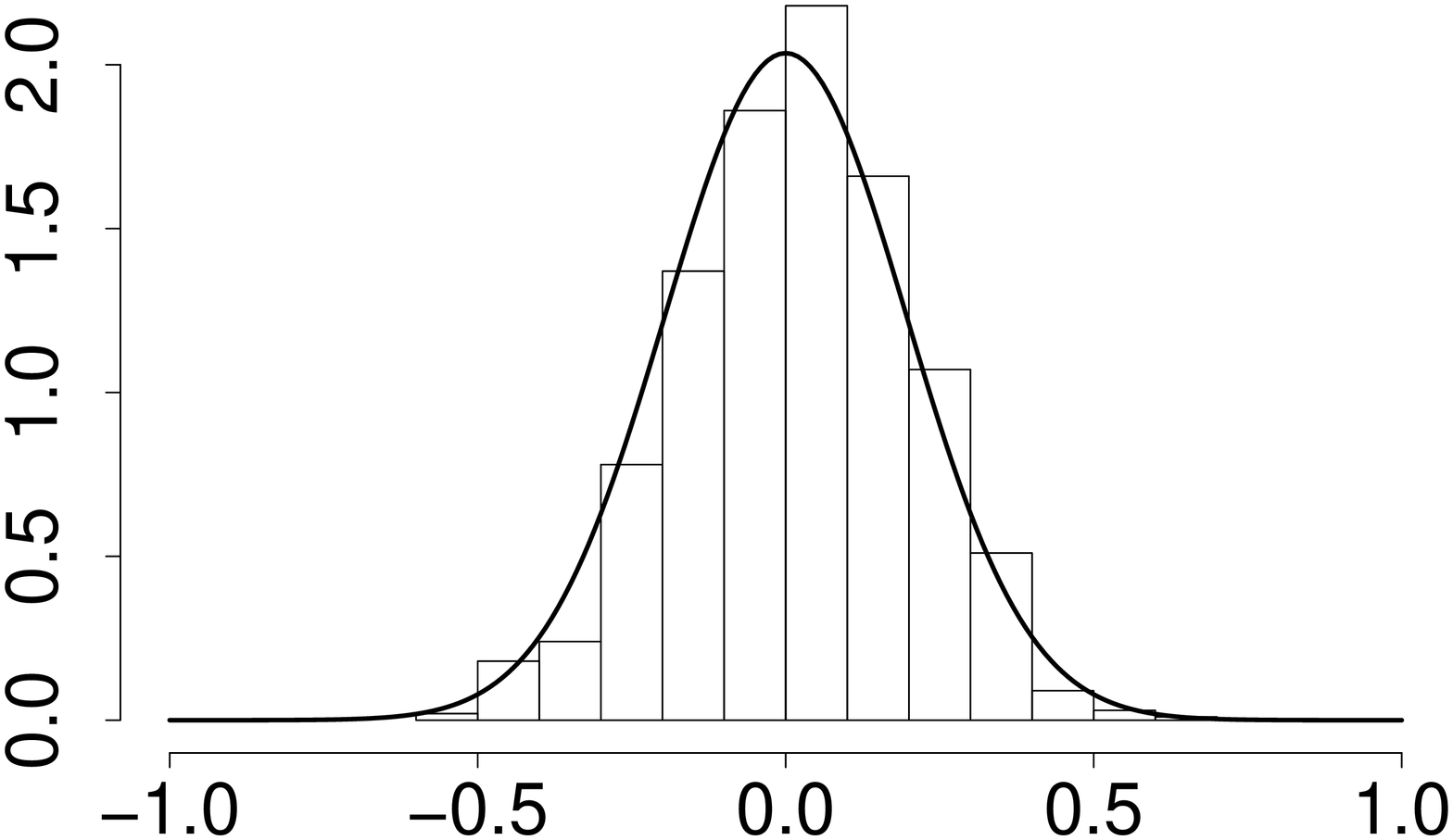}}&
\rotatebox{270}{\includegraphics[width=0.2\linewidth]{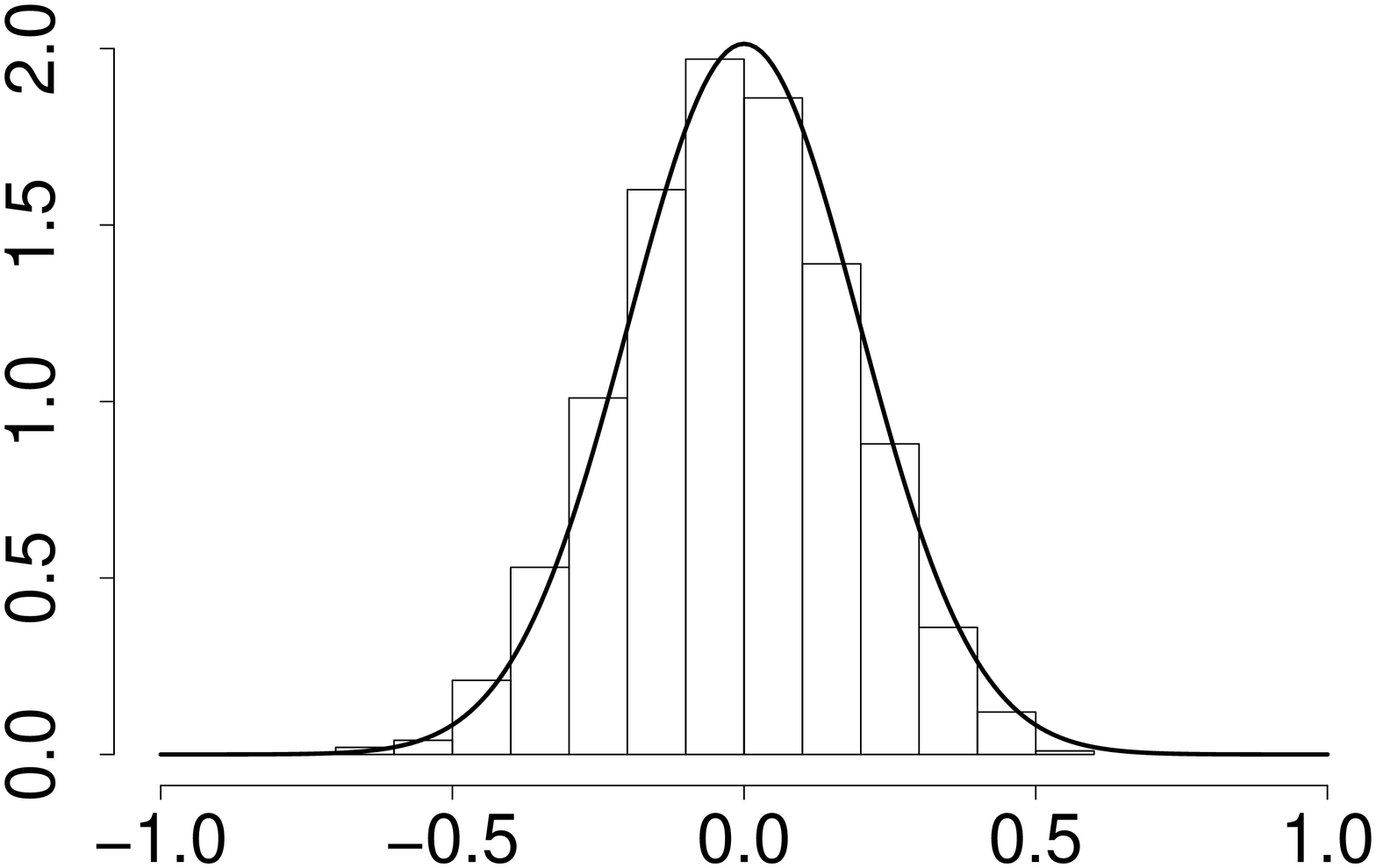}}&
\rotatebox{270}{\includegraphics[width=0.2\linewidth]{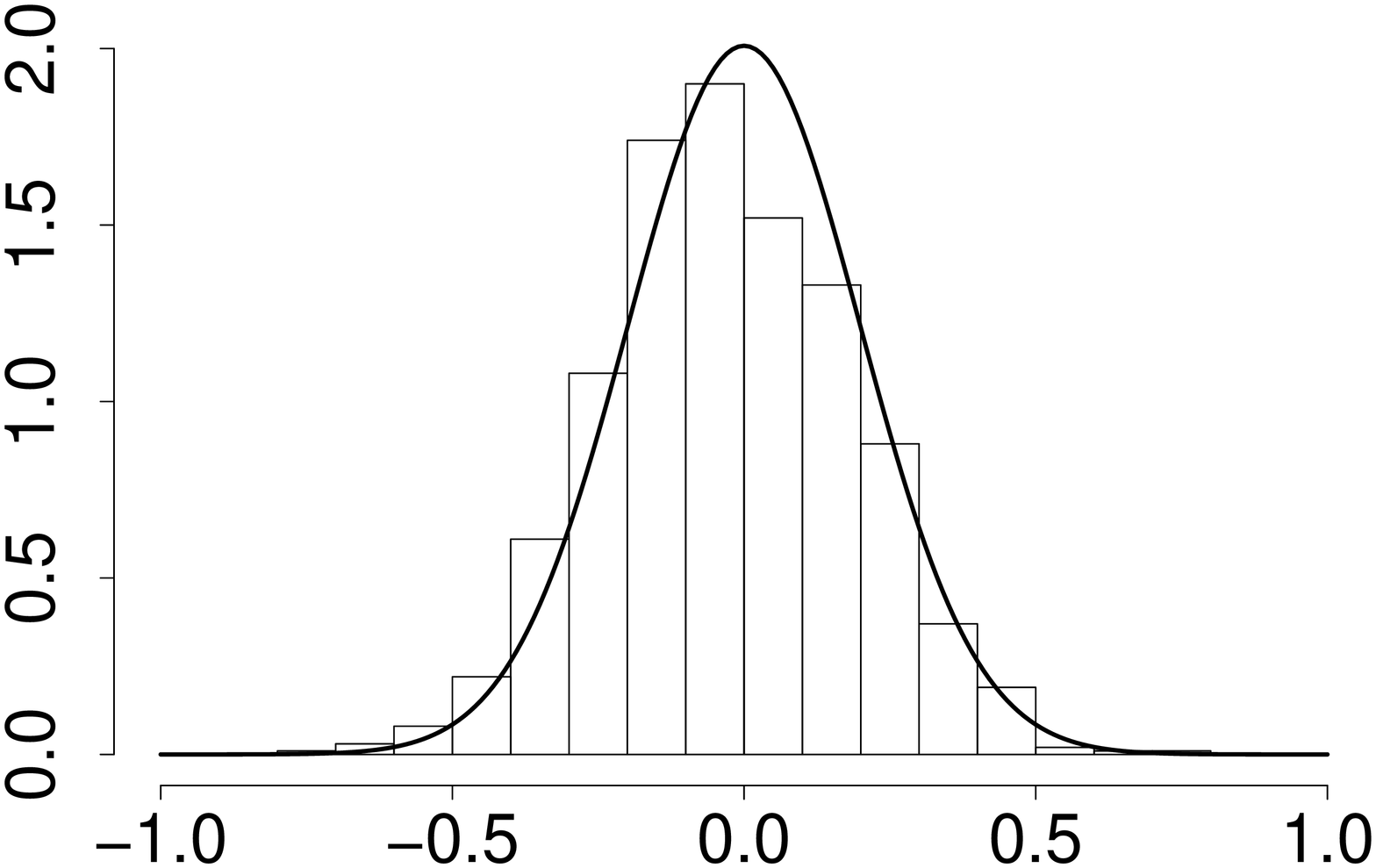}}
\end{tabular}
\end{center}
\caption{\sl Results for the Gaussian drawable vine with $\rho=0$$\cdot$$5$ at levels 2, 3 and 4 (in columns 1, 2 and 3, respectively). The first row shows the means of the samples of the absolute difference between $\CC_{ij|v,n}(0$$\cdot$$3,0$$\cdot$$7)$ and expansion \eqref{eq:emppaircopproc}, for $n$ from $10^{2}$ to $10^{5}$, on log-log scale (the original values are on the axes). The last two rows display normal QQ-plots and histograms of the samples of $\CC_{ij|v,n}(0$$\cdot$$4,0$$\cdot$$2)$ with $n=1,000$, respectively, the latter superposed by the limiting probability density functions from \eqref{eq:asnorm}.}
\label{fig:Eq_16_19_20}
\end{figure}

According to \eqref{eq:emppaircopproc}, the absolute difference between the
left and right hand sides should decrease and eventually vanish as $n$ increases. 
The top row of Figure \ref{fig:Eq_16_19_20} shows the mean of these differences 
in the point $(0$$\cdot$$3,0$$\cdot$$7)$, over the simulations from the Gaussian 
drawable vine with $\rho=0$$\cdot$$5$, for growing $n$, on log-log scale. Indeed, 
these decrease, though rather slowly. The rate of convergence appears to be 
approximately of the same order as for the ordinary empirical copula process, 
namely $n^{-1/4}$, which is to be expected.

Furthermore, we have tested the limiting distribution \eqref{eq:asnorm} of 
$\CC_{ij|v,n}(u_i, u_j)$ with variance \eqref{eq:asvar}, using the 
Kolmogorov-Smirnov goodness-of-fit test. Table \ref{tab:gof} shows the
corresponding p-values in the three points $(0$$\cdot$$1,0$$\cdot$$3)$, 
$(0$$\cdot$$4,0$$\cdot$$2)$ and $(0$$\cdot$$7,0$$\cdot$$8)$ for a selection of 
models and levels, with $\rho=0$$\cdot$$5$ and $n=1,000$. The consistently high
p-values indicate that the assumed distribution fits the samples well.
This is confirmed by normal QQ-plots and histograms of the samples, 
superposed by the asymptotic probability density functions. These are displayed
in the lower two rows of Figure \ref{fig:Eq_16_19_20} for the Gaussian drawable 
vine with $n=1,000$. Examples with other structures and copulae may be found in 
the supplement.

\begin{table}
\centering
\def~{\hphantom{0}}
\begin{tabular}{llcccc}
 \textsc{Structure} & \textsc{Copula} & \textsc{Level} & \multicolumn{3}{c}{\textsc{p-value}}\\
 & & & (0$\cdot$1,0$\cdot$3) & (0$\cdot$4,0$\cdot$2) & (0$\cdot$7,0$\cdot$8)\\[5pt]
\multirow{5}{\LL}{Drawable} & \multirow{3}{\LLLL}{Gaussian} & 2 & 0$\cdot$54 & 0$\cdot$48 & 0$\cdot$67\\
 & & 3 & 0$\cdot$40 & 0$\cdot$70 & 0$\cdot$69\\
 & & 4 & 0$\cdot$26 & 0$\cdot$44 & 0$\cdot$36\\
 & Student's t & 2 & 0$\cdot$47 & 0$\cdot$65 & 0$\cdot$51\\
 & Gumbel & 2 & 0$\cdot$37 & 0$\cdot$91 & 0$\cdot$45\\[1ex]
 Canonical & Gaussian & 3 & 0$\cdot$37 & 0$\cdot$45 & 0$\cdot$44\\[1ex]
 Regular & Gaussian & 4 & 0$\cdot$35 & 0$\cdot$61 & 0$\cdot$41
\end{tabular}
\caption{\sl P-values from the Kolmogorov-Smirnov goodness-of-fit tests on the simulated processes $\CC_{ij|v,n}(u_i, u_j)$ with $n=1,000$, in the points (0$\cdot$1,0$\cdot$3), (0$\cdot$4,0$\cdot$2) and (0$\cdot$7,0$\cdot$8), for different vine structures, copula models and levels}
\label{tab:gof}
\end{table}

As seen from \eqref{eq:condground} and \eqref{eq:condhigh}, the estimators 
depend on a bandwidth parameter $h_{n}$. This bandwidth should obviously be 
proportional to some power of $n^{-1}$, at least $n^{-1/2}$ to guarantee 
consistent estimators. Viewing $\hat{C}_{ij,n}^{[2]}(u_{i}, u_{j})$ and 
$\hat{C}_{ij|v,n}^{[2]}(u_{i}, u_{j})$ as predictions of the conditional 
expectations $\E(I(U_{i} \leq u_{i}) | U_{j} = u_{j}))$ and 
$\E(I(U_{i|v} \leq u_{i}) | U_{j|v} = u_{j}))$, respectively, one could 
construct a cross-validation procedure for bandwidth selection. However,
this is non-trivial since the observations are indicator functions, the choice 
of points $(u_{i},u_{j})$ is not obvious and there are up to $(d-1)(d-2)$ 
predictors to evaluate.

In order to investigate the influence of the bandwidth on the estimators,
we have repeated the simulation of the Gaussian drawable vine with each of 
the three bandwidths $h_{n}=n^{-\frac{1}{5}}$, $n^{-\frac{1}{4}}$ and 
$0$$\cdot$$5n^{-\frac{1}{3}}$. The first of these is proportional to Silverman's 
rule. The other two are undersmoothing alternatives. As mentioned earlier, 
the optimal choice of bandwidth does not depend on the margins, but only on the 
dependence structure. For $\rho=0$$\cdot$$2$, the estimators behave well for all 
three bandwidths, but for higher dependencies, i.e. $\rho=0$$\cdot$$5$ and 
$0$$\cdot$$8$, $h_{n}=0$$\cdot$$5n^{-\frac{1}{3}}$ is without a doubt the only 
sensible choice. We have therefore used that bandwidth throughout the paper. 
Intuitively, it makes sense to undersmooth the estimators, i.e. to minimize the 
bias at the expense of the variance. The pointwise estimates of the conditional 
distributions may then differ considerably from the true values, but the copula 
estimator will average out these discrepancies. 

It remains to test the proposed resampling procedure. We have simulated from the 
Gaussian, Student's t and Gumbel drawable vines from before, in dimension $d=4$, 
with $\rho=0$$\cdot$$5$ and $n=1,000$. For each sample, we have generated $B=1,000$ 
multiplier bootstrap estimates of the confidence interval \eqref{eq:cimultboot} of 
the top level copula $C_{14|23}$, evaluated in $(0$$\cdot$$4,0$$\cdot$$2)$, with 
$\alpha=\{0$$\cdot$$1,0$$\cdot$$05,0$$\cdot$$01\}$, using both approaches, as well 
as plug-in estimates, based on $\hat{\sigma}_{ij|v}^{2}$. 

As suggested earlier, parametric estimators are more efficient when the model is 
correctly specified, or at least close to the truth. However, we believe that 
the empirical estimator is more robust. Therefore, we have computed corresponding 
percentile confidence intervals based on parametric bootstrap, assuming both correct 
and incorrect copula families in the lower two levels, but always the true family 
for the copula of interest. More specifically, we estimated the intervals for the 
Gaussian model, assuming first the true model, and then Gumbel copulae in the first 
two levels. We repeated this for the Student's t model with Gumbel copulae, and for 
the Gumbel model with Gaussian copulae. The estimator we have used is the stepwise 
semiparametric estimator; see for instance \cite{haff:2012}. 

Table \ref{tab:resampling} shows the confidence intervals' average length and 
actual coverage, i.e. the fraction of intervals that contain the true value of 
$C_{12|34}(0$$\cdot$$4,0$$\cdot$$2)$. The ones based on the multiplier bootstrap 
percentiles are shorter than the symmetric ones. Further, the plug-in estimator 
$\hat{\sigma}_{14|23}^{2}$ is surprisingly good. Of course, all these intervals 
are longer than the ones obtained with the parametric estimator. Moreover, their 
actual coverage is consistently higher than the nominal one. However, the 
misspecified models produce intervals with substantially lower coverage than the 
chosen confidence levels. Hence, tests based on the empirical estimator are expected
to be conservative, and thus less powerful than parametric equivalents, but on the
other hand more robust towards misspecifications in lower levels. Another advantage 
of the multiplier bootstrap scheme is that it is much faster than the parametric one, 
especially for the Student's t model. Also note that for this particular model, 
the parametric intervals made under the true model assumptions have smaller coverage 
than the nominal one, which probably means that $B=1,000$ is insufficient in this case. 
Naturally, the misspecifications in the above experiments are not very realistic, 
but merely meant as an illustration of how errors propagate from level to level. 
In practice, one should be able to choose reasonably well at least at the first 
level. 

We repeated the above simulations for the Gaussian model with $B=500$ and with 
$n=10,000$. The results were as expected. When $n$ increases, the interval lengths 
obviously decrease, whereas the actual coverage becomes more varying for smaller $B$. 
We therefore use $n=B=1,000$ in the remaining sections. 

\begin{table}
\centering
\def~{\hphantom{0}}
\begin{tabular}{@{}llccccc}%\small
\textsc{Model} & \textsc{$\alpha$} & \multicolumn{3}{c}{\textsc{Non-parametric}} & \multicolumn{2}{c}{\textsc{Parametric}}\\
 & & Percentile & Symmetric & Plug-in & Correct & Incorrect\\[5pt]
\multirow{3}{\LLLL}{Gaussian} & 0$\cdot$1 & 0$\cdot$94~(2$\cdot$1) & 0$\cdot$94~(2$\cdot$1) & 0$\cdot$94~(2$\cdot$1) & 0$\cdot$90~(1$\cdot$1) & 0$\cdot$86~(1$\cdot$1)\\
 & 0$\cdot$05 & 0$\cdot$97~(2$\cdot$5) & 0$\cdot$98~(2$\cdot$5) & 0$\cdot$98~(2$\cdot$5) & 0$\cdot$95~(1$\cdot$3) & 0$\cdot$92~(1$\cdot$3)\\
 & 0$\cdot$01 & 0$\cdot$99~(3$\cdot$2) & 0$\cdot$99~(3$\cdot$2) & 0$\cdot$99~(3$\cdot$3) & 0$\cdot$99~(1$\cdot$7) & 0$\cdot$97~(1$\cdot$7)\\
\\
 \multirow{3}{\LLLL}{Student's t} & 0$\cdot$1 & 0$\cdot$91~(2$\cdot$1) & 0$\cdot$91~(2$\cdot$1) & 0$\cdot$91~(2$\cdot$1) & 0$\cdot$89~(1$\cdot$2) & 0$\cdot$00~(1$\cdot$1)\\
 & 0$\cdot$05 & 0$\cdot$95~(2$\cdot$5) & 0$\cdot$95~(2$\cdot$5) & 0$\cdot$95~(2$\cdot$5) & 0$\cdot$94~(1$\cdot$4) & 0$\cdot$00~(1$\cdot$4)\\
 & 0$\cdot$01 & 0$\cdot$99~(3$\cdot$2) & 0$\cdot$99~(3$\cdot$3) & 0$\cdot$99~(3$\cdot$3) & 0$\cdot$98~(1$\cdot$9) & 0$\cdot$00~(1$\cdot$8)\\
\\
 \multirow{3}{\LLLL}{Gumbel} & 0$\cdot$1 & 0$\cdot$91~(2$\cdot$0) & 0$\cdot$91~(2$\cdot$0) & 0$\cdot$91~(2$\cdot$0) & 0$\cdot$91~(1$\cdot$0) & 0$\cdot$68~(1$\cdot$0)\\
 & 0$\cdot$05 & 0$\cdot$96~(2$\cdot$4) & 0$\cdot$96~(2$\cdot$4) & 0$\cdot$96~(2$\cdot$4) & 0$\cdot$96~(1$\cdot$2) & 0$\cdot$79~(1$\cdot$2)\\ 
 & 0$\cdot$01 & 0$\cdot$99~(3$\cdot$1) & 0$\cdot$99~(3$\cdot$1) & 0$\cdot$99~(3$\cdot$1) & 0$\cdot$99~(1$\cdot$6) & 0$\cdot$91~(1$\cdot$5)\\ 
\end{tabular}
\caption{\sl Coverage of the estimated confidence intervals for $C_{14|23}(0$$\cdot$$4,0$$\cdot$$2)$ in the Gaussian, Student's t and Gumbel models with $n=1,000$. The average lengths, in parentheses, are multiplied by $10^2$.}
\label{tab:resampling}
\end{table}

\section{Methods derived from the empirical pair-copula}

\subsection{Estimating the conditional Spearman correlation}
\label{subsec:sprho}

The Spearman correlation $\rho_S$ of the bivariate copula $C$ of  a random
pair $(U, V)$ with uniform $(0, 1)$ margins can be expressed as
\[
  \rho_S(C) = 12 \int_{[0, 1]^{2}} C(u, v) \, \diff (u, v) - 3 = \corr(U, V).
\]
Similarly, $\rho_S( C_{ij|v} )$ is a measure of association between $X_i$ and $X_j$, 
conditionally on $X_v$. This quantity can be estimated by the plug-in estimator 
$\rho_S(\hat{C}_{ij|v,n}) = 12 \int \hat{C}_{ij|v,n} - 3$, which is approximately 
equal to the sample correlation of the pairs $(\hat{U}_{i|v,tn},\hat{U}_{j|v,tn})$, 
for $t = 1, \ldots, n$. The expansion of the empirical pair-copula process in 
\eqref{eq:emppaircopproc} implies that
\begin{equation}
\label{eq:asnormsprho}
  n^{1/2} \{ \rho_S( \hat{C}_{ij|v,n} ) - \rho_S( C_{ij|v} ) \} \dto 12 \int_{[0, 1]^2} \CC_{ij|v}(u_i, u_j) \, \diff(u_i, u_j),
\end{equation}
where $\CC_{ij|v}$, the conjectured large-sample limit of $\CC_{ij|v,n}$, is a 
centred Gaussian process on $[0, 1]^2$ with covariance function determined by the 
right-hand side of \eqref{eq:emppaircopproc}. The limiting random variable in 
\eqref{eq:asnormsprho} is a centred normal random variable with variance
\[
  \sigma_{\rho}^{2} = 144 \int_{[0, 1]^{2}} \int_{[0, 1]^{2}} \cov \{ \CC_{ij|v}(u_i, u_j), \, \CC_{ij|v}(u_i', u_j') \} \, \diff(u_i, u_j) \, \diff(u_i', u_j').
\]
This variance can be estimated either by a plug-in estimator or via the multiplier 
resampling scheme described in Section~\ref{s:resampling}. The latter procedure 
consists in resampling $\CC_{ij|v,n}$ and integrating it either by numerical or Monte 
Carlo integration over $[0, 1]^{2}$. One may then estimate $\sigma_{\rho}^{2}$ by the 
sample variance of the $B$ resamples of $12 \int \CC_{ij|v}$. Further, confidence 
intervals for $\rho_S(C_{ij|v,n})$ can be constructed either via the normal 
approximation with estimated variance $\sigma_{\rho}^{2}$ or by using resample 
percentiles.

In order to verify \eqref{eq:asnormsprho}, we have simulated from the same 
four-dimensional models as in the last part of Section \ref{subsec:proofbysimul}, 
computing \eqref{eq:asnormsprho} for each of the $1,000$ samples. The p-values 
from the Kolmogorov-Smirnov tests are $0$$\cdot$$89$, $0$$\cdot$$92$ and $0$$\cdot$$68$, 
respectively, for the three models, which clearly agrees with the conjecture. 
Normal QQ-plots and histograms are shown in the supplement.

Moreover, we have tested the suggested resampling scheme for $12 \int \CC_{14|23}$
in the same way as in Section \ref{subsec:proofbysimul}. The corresponding results,
shown in the supplement, are very similar. The confidence intervals based on the 
empirical estimator are longer and have larger actual coverage than the parametric 
equivalents, whereas the latter are non-robust towards misspecifications 
in lower levels. Once more, the intervals based on the multiplier bootstrap 
percentiles appear to be the best of the empirical ones. The plug-in estimator 
of the variance $\sigma_{\rho}^{2}$ is also rather good, but computationally 
much slower than the multiplier bootstrap. 

\subsection{Vine structure selection}
\label{subsec:vine}

Selecting the structure of a pair-copula construction consists in choosing 
which variables to associate with a pair-copula at each level. As the model 
uncertainty increases with the level, the state of the art is to try to 
capture as much of the dependence as possible in the lower levels of the 
structure. \cite{Vines} propose ordering the variables of a drawable vine in 
the way that maximizes the tail dependencies at the ground level, while 
\cite{dissmann11} suggest a model selection algorithm for more general regular
vines, that maximizes the sum of absolute values of Kendall's $\tau$ coefficients 
at each level. Both these schemes require the simultaneous choice and estimation 
of parametric copulae. At the ground level, the latter algorithm only uses the 
sample Kendall's $\tau$s, and therefore does not call for assumptions about 
the pair-copulae. However, from the second level on, the $\tau$s involve the 
unobserved variables $F_{i|v}(X_i|X_v)$, that are estimated parametrically from 
the copulae in the previous level via \eqref{eq:Fijv}. Inadequate choices of 
copulae may thus influence the structure selection at the higher levels. 

We propose a more robust model selection scheme, based on our nonparametric 
estimate of the Spearman correlation $\rho_S$. \par\medskip

%\begin{algorithm}
\begin{compactenum}
  \item Compute the ground level normalized ranks $\hat{U}_{itn}$ ($t=1,\ldots,n)$.
  \item Compute $\rho_S(\hat{C}_{ij,n})$ for all pairs $\{i, j\} \subset \{1, \ldots, d\}$ such that $i < j$.
  \item Select the spanning tree $T_1$ on $\{1, \ldots, d\}$ that maximizes $\sum_{\{i, j\} \in T_1} |\rho_S(\hat{C}_{ij,n})|$.
  \item Estimate $\hat{U}_{i|j,tn}$ and $\hat{U}_{j|i,tn}$ for all selected pairs $\{i, j\}$, using \eqref{eq:condground}.
  \item For levels $\ell = 2, \ldots, d-1$:
    \begin{compactenum}
      \item Compute $\rho_S(\hat{C}_{ij|v,n})$ for all possible pairs $\{i, j\}$. 
      \item Select the spanning tree $T_\ell$ that maximizes $\sum_{\{i, j\} \in T_\ell }|\rho_S(\hat{C}_{ij|v,n})|$.
      \item Estimate $\hat{U}_{i|j \cup v,tn}$ and $\hat{U}_{j|i \cup v,tn}$ for all selected pairs $\{i, j\}$, using \eqref{eq:condhigh}.
    \end{compactenum}
\end{compactenum} \par \medskip

The above algorithm is strongly inspired by \cite{dissmann11}, who also explain 
the concept of possible pairs. We merely estimate the copulae and conditional 
distributions nonparametrically rather than parametrically and use Spearman's 
$\rho_S$ instead of Kendall's $\tau$. The substitution of dependence measures 
should not influence the results too much. When the model is well specified, one 
would therefore expect the two algorithms to select virtually the same structure.

The algorithm is put into practice in Section~\ref{subsec:finance}, where it is 
found to impose quite a reasonable structure on a set of financial variables.

\subsection{Testing for conditional independence}
\label{subsec:indtest}

The number of parameters in a pair-copula construction grows rapidly with increasing 
dimension $d$. Identifying independence copulae in the structure is one way of 
reducing this number. One may therefore add tests for conditional independence as 
a step in the model selection algorithm of Section~\ref{subsec:vine}.

In case $C_{ij|v}$ is the independence copula, equation~\eqref{eq:emppaircopproc} 
implies that the asymptotic distribution of $\hat{C}_{ij|v,n}$ is the same as the one 
of the bivariate empirical copula under independence. In other words, the random vectors 
$(\hat{U}_{i|v,tn}, \hat{U}_{j|v,tn})$ for $t \in \{1, \ldots, n\}$ behave in distribution 
as the sample of bivariate normalized ranks from a random sample of a bivariate distribution 
with independent components. Therefore, rank-based tests for independence can be applied 
without adjustment of the critical values.

We propose to test the null hypothesis of conditional independence of $X_i$ and $X_j$, 
given $X_v$, by the Cram{\'e}r-von Mises test statistic
\[
  \int_{[0, 1]^2} \CC_{ij|v,n}^2(u_i, u_j) \, \diff \hat{C}_{ij|v,n}(u_i, u_j)
  = \frac{1}{n} \sum_{t=1}^n \CC_{ij|v,n}^2( \hat{U}_{i|v,tn}, \hat{U}_{j|v,tn} ),
\]
where $\CC_{ij|v,n}(u_i, u_j) = n^{1/2} \{ \hat{C}_{ij|v,n}(u_i, u_j) - u_i u_j \}$ is the 
empirical pair-copula process under the null hypothesis of conditional independence, that 
is, $C_{ij|v}(u_i, u_j) = u_i u_j$ for all $(u_i, u_j) \in [0, 1]^2$. Under the null 
hypothesis, the limit distribution of the test statistic is distribution free and is given 
by $\int_{[0, 1]^2} \CC^2(u, v) \, \diff(u, v)$, where $\CC$ is the limiting empirical copula 
process under independence. Critical values of the test statistic can be obtained by Monte 
Carlo estimation based on random samples from a distribution with independent components.

Once more, we have compared our test with parametric equivalents, based 
on parametric bootstrap, on the four-dimensional models from Section 
\ref{subsec:proofbysimul}, but with the top level copula 
$C_{14|23}(u_{1},u_{4})=u_{1}u_{4}$. Table \ref{tab:indep_test} shows the rejection 
rates at levels $\alpha=\{0$$\cdot$$1,0$$\cdot$$05,0$$\cdot$$01\}$. Again, the tests 
based on the empirical estimator appear to be conservative, that is, the rejection 
rates are consistently lower than the specified levels. As anticipated, the parametric 
tests are more powerful under correct model assumptions, but the rejection rates are 
slightly too high for the Student's t model, which seems to require a higher $B$. 
Moreover, the rejection rates are too high under incorrect model assumptions, which 
demonstrates these tests' lack of robustness.

\subsection{Goodness-of-fit testing}
\label{subsec:gof}

In the parametric case, model selection consists in choosing not only the
structure, as described in Section \ref{subsec:sprho}, but also the 
families of the $d(d-1)/2$ copulae. Goodness-of-fit tests can help to 
assess whether the selected model represents the dependence structure well. 
At the ground level, one may simply apply the standard tests, for instance 
the ones studied in \cite{genest09}. From the second level on, it becomes
more complicated, since the copula arguments are themselves unknown conditional 
distributions, derived from a cascade of pair-copulae at lower levels.

Following the reasoning of Section \ref{subsec:indtest}, we propose a
Cram{\'e}r-von Mises good\-ness-of-fit test, more specifically, the test
proposed by \cite{genest08}, replacing the normalized ranks by our
non-parametric estimators of the conditional distributions. Critical
values may then be obtained by the bootstrap procedure they describe, again 
substituting the normalized ranks by our estimators.

Testing this procedure on the top level copula of the four-dimensional Gumbel
model from Section \ref{subsec:proofbysimul} with  $n=1,000$, we obtained rejection 
rates of $0$$\cdot$$098$, $0$$\cdot$$042$ and $0$$\cdot$$0044$ for the null hypothesis 
that it is a Gumbel copula at levels $0$$\cdot$$1$, $0$$\cdot$$05$ and $0$$\cdot$$01$, 
respectively. For the hypotheses that it is Student's t and Gaussian, the corresponding 
rates were $0$$\cdot$$90$, $0$$\cdot$$83$, $0$$\cdot$$60$ and $0$$\cdot$$91$, 
$0$$\cdot$$84$, $0$$\cdot$$62$. Hence the former are clearly rejected, while 
the true model, Gumbel, is not, as it should be. 

\begin{table}
\centering
\def~{\hphantom{0}}
\begin{tabular}{lllll}
\textsc{Model} & & \textsc{Non-parametric} & \multicolumn{2}{c}{\textsc{Parametric}}\\
 & $\alpha$ & & Correct & Incorrect\\[5pt]
\multirow{3}{\LLLL}{Gaussian} & 0$\cdot$1 & 0$\cdot$099 & 0$\cdot$10 & 0$\cdot$22\\
 & 0$\cdot$05 & 0$\cdot$049 & 0$\cdot$050 & 0$\cdot$14\\
 & 0$\cdot$01 & 0$\cdot$0096 & 0$\cdot$010 & 0$\cdot$034\\
\\
\multirow{3}{\LLLL}{Student's t} & 0$\cdot$1 & 0$\cdot$097 & 0$\cdot$11 & 0$\cdot$16\\
 & 0$\cdot$05 & 0$\cdot$047 & 0$\cdot$068 & 0$\cdot$082\\
 & 0$\cdot$01 & 0$\cdot$0093 & 0$\cdot$022 & 0$\cdot$042\\
\\
\multirow{3}{\LLLL}{Gumbel} & 0$\cdot$1 & 0$\cdot$094 & 0$\cdot$099 & 0$\cdot$23\\
 & 0$\cdot$05 & 0$\cdot$044 & 0$\cdot$048 & 0$\cdot$14\\
 & 0$\cdot$01 & 0$\cdot$0088 & 0$\cdot$0090 & 0$\cdot$048\\
\end{tabular}
\caption{\sl Rejection rates from the Cram{\'e}r-von Mises tests for conditional independence at the third level of the Gaussian, Student's t and Gumbel models with $n=1,000$}
\label{tab:indep_test}
\end{table}

\section{Data examples}
\label{s:data}

\subsection{Financial data set}
\label{subsec:finance}

The financial data set consists of nine Norwegian and international daily 
price series from March 25th, 2003, to March 26th, 2006, which corresponds 
to 1107 observations. These include the Norwegian 5- and 6-year Swap Rates 
(NI5 and NI6), the 5-year German Government Rate (GI5), the BRIX Norwegian 
Bond Index (NB) and ST2X Government Bond Index (MM), the WGBI Citigroup World 
Government Bond Index (IB), the OSEBX Oslo Stock Exchange Main Index (NS), the 
MSCI Morgan Stanley World Index (IS) and the Standard \& Poor Hedge Fund Index 
(HF). This is a subset of the 19 variables, analyzed in 
\cite{brechmann:czado:aas:2011}, which represent the market portfolio of one 
of the largest Norwegian financial institutions. We have followed their example, 
and filtrated each of the series with an appropriate time series model to remove
the temporal dependence. Subsequently, we have modelled the standardized 
residuals with a regular vine. 

\begin{figure}
%\begin{center}
\centering
\includegraphics[height=0.3\textheight]{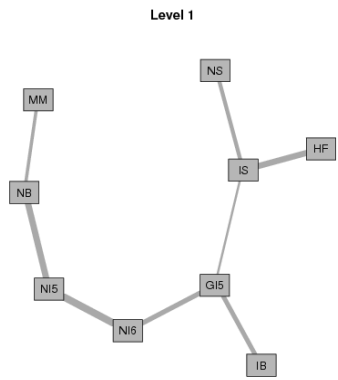}
\caption{\sl Ground level of the regular vine selected for the financial data in Section~\ref{subsec:finance}. The thickness of the edges is determined by the absolute value of the corresponding Spearman rank correlations.}
\label{fig:findataTree}
%\end{center}
\end{figure}

We selected the vine structure, first with the method proposed in 
Section~\ref{subsec:vine} and then with the method of \cite{dissmann11}. The 
two selected structures were actually identical, which is reassuring. The dependence 
in the ground level appears to be very strong, with Spearman rank correlations that 
are large in absolute value. In the remaining levels, the Spearman correlations are 
considerably smaller, and only 9 out of 28 copulae were significantly different from 
the independence copula at level 0$\cdot$05, according to the test from 
Section~\ref{subsec:indtest}. Hence, most of the dependence has been captured in 
the ground level, shown in Figure~\ref{fig:findataTree}, which was the aim.

The collection of pairs selected by the algorithm at this level is quite reasonable. 
The three stock indices and the three interest rates are grouped together, whereas 
the Norwegian bond indices are dependent on the international bond index via the 
interest rates. 

\subsection{Precipitation data set}
\label{subsec:precipitation}

The precipitation set is composed of daily recordings from January 1st, 1990, 
to December 31st, 2006, at five different meteorological stations in Norway:
Vestby, Ski, L{\o}renskog, Nannestad and Hurdal. This data set was used in both 
\cite{bergaas09} and \cite{haff:2012}. As in those papers, we have modelled only 
the positive precipitation, discarding all observations for which at least one 
of the stations has recorded zero precipitation. The remaining 2013 observations
appear to be fairly independent in time. We model these with a drawable vine,
ordering the stations according to geography. The model is quite natural since the stations are located almost on a straight line,
%in the above given order
from Vestby in the South to Hurdal in the North; see the map in
the supplement. The parametric model used for comparison is the one from 
\cite{haff:2012}, with Gumbel copulae at the ground level and subsequently 
Gaussian ones.

Since rain showers tend to be rather local, one would expect the dependence to 
be strongest between the closest stations, and decrease with the level, possibly 
even down to conditional independence. Therefore we have tested the second, 
third and fourth level copulae for conditional independence, both with the 
non-parametric test from Section~\ref{subsec:indtest} and equivalent parametric 
tests. The Spearman rank correlations at the ground level range from 0$\cdot$82 to 
0$\cdot$94, indicating a strong positive dependence. At the second level, they are 
considerably lower, but the hypothesis of conditional independence is rejected for 
all copulae, by both tests, which actually agree in the last two levels as well. 
The conditional copula of the measurements from Vestby and Nannestad, given the 
two stations in between, is also significantly different from independence. This 
is not true for Ski and Hurdal, conditioning on L{\o}renskog and Nannestad, and 
neither for the top level copula, linking Vestby and Hurdal, conditionally on the 
three stations in between.
\section*{Acknowledgement}
J. Segers gratefully acknowledges funding from the Belgian Science 
Policy and the Acad\'emie universitaire `Louvain'. This work has also been 
funded by Statistics for Innovation, (sfi)$^2$.

\bibliographystyle{chicago} 
\bibliography{biblio,references} 

\clearpage
\appendix

\section{Supplement: Extra figures and tables}

\mbox{}

\begin{figure}[h]
\begin{center}
\begin{tabular}{c@{\quad}c}
\rotatebox{270}{\includegraphics[width=0.28\linewidth,height=0.22\linewidth]{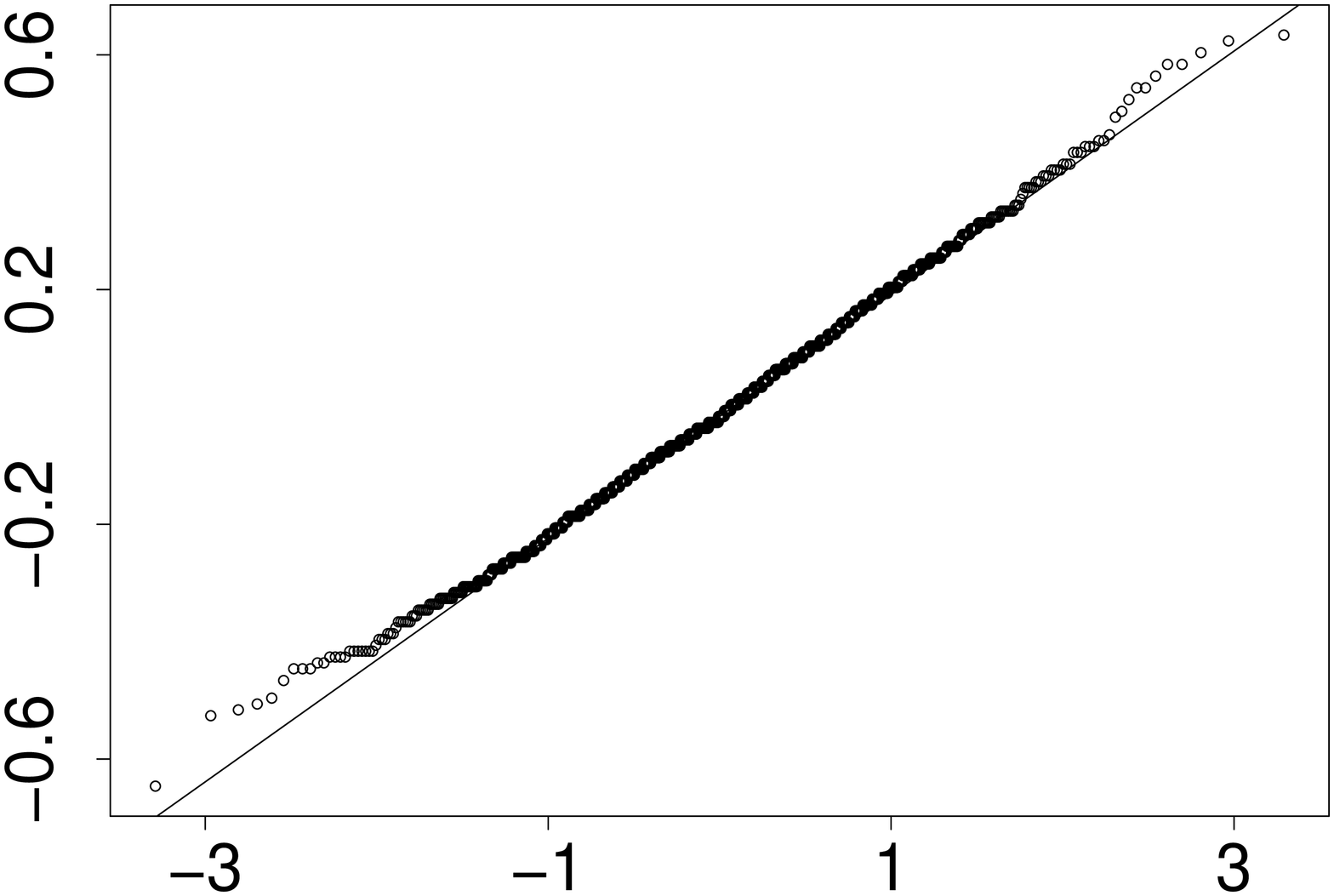}}&
\rotatebox{270}{\includegraphics[width=0.28\linewidth,height=0.22\linewidth]{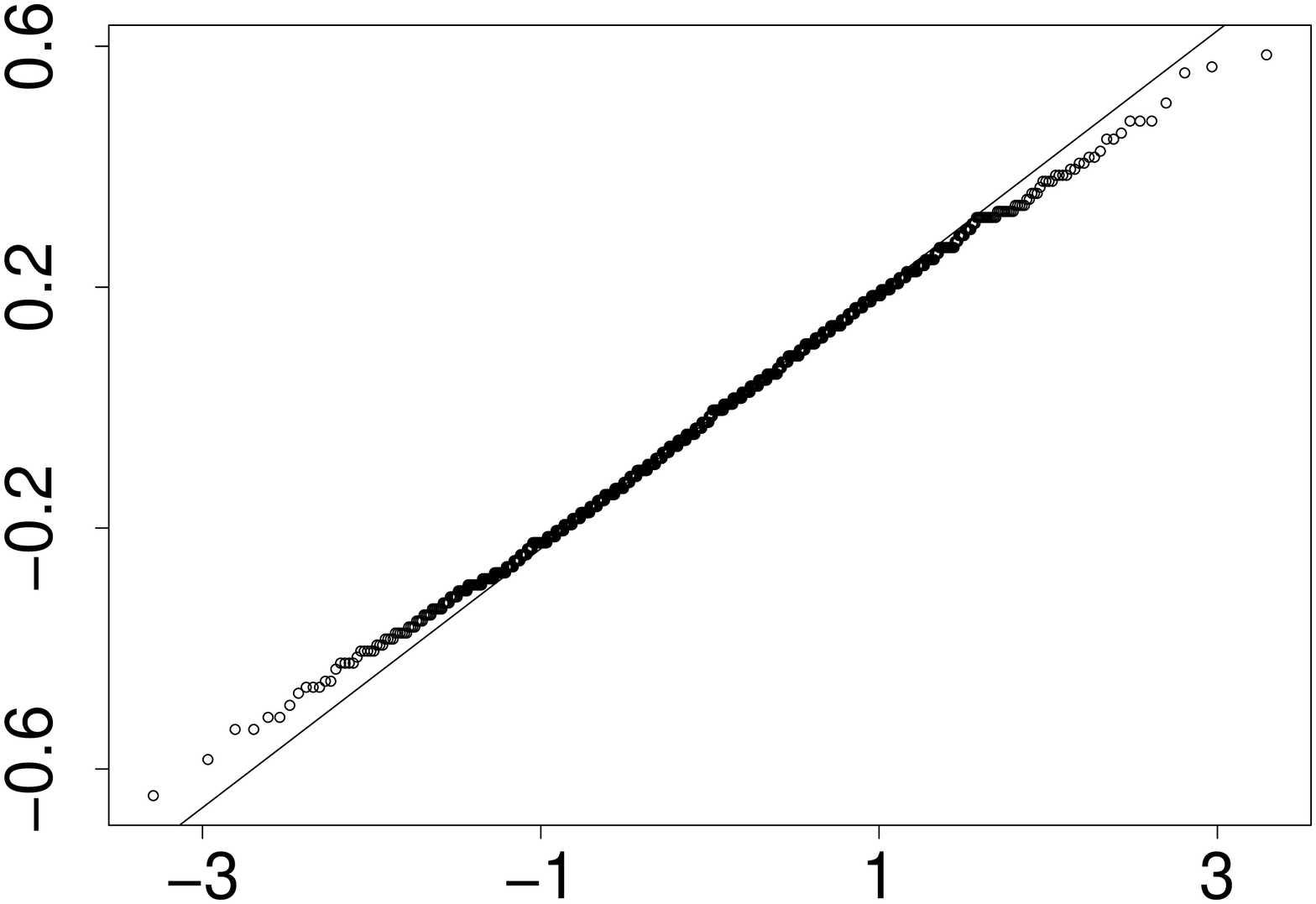}}\\
\rotatebox{270}{\includegraphics[width=0.28\linewidth,height=0.22\linewidth]{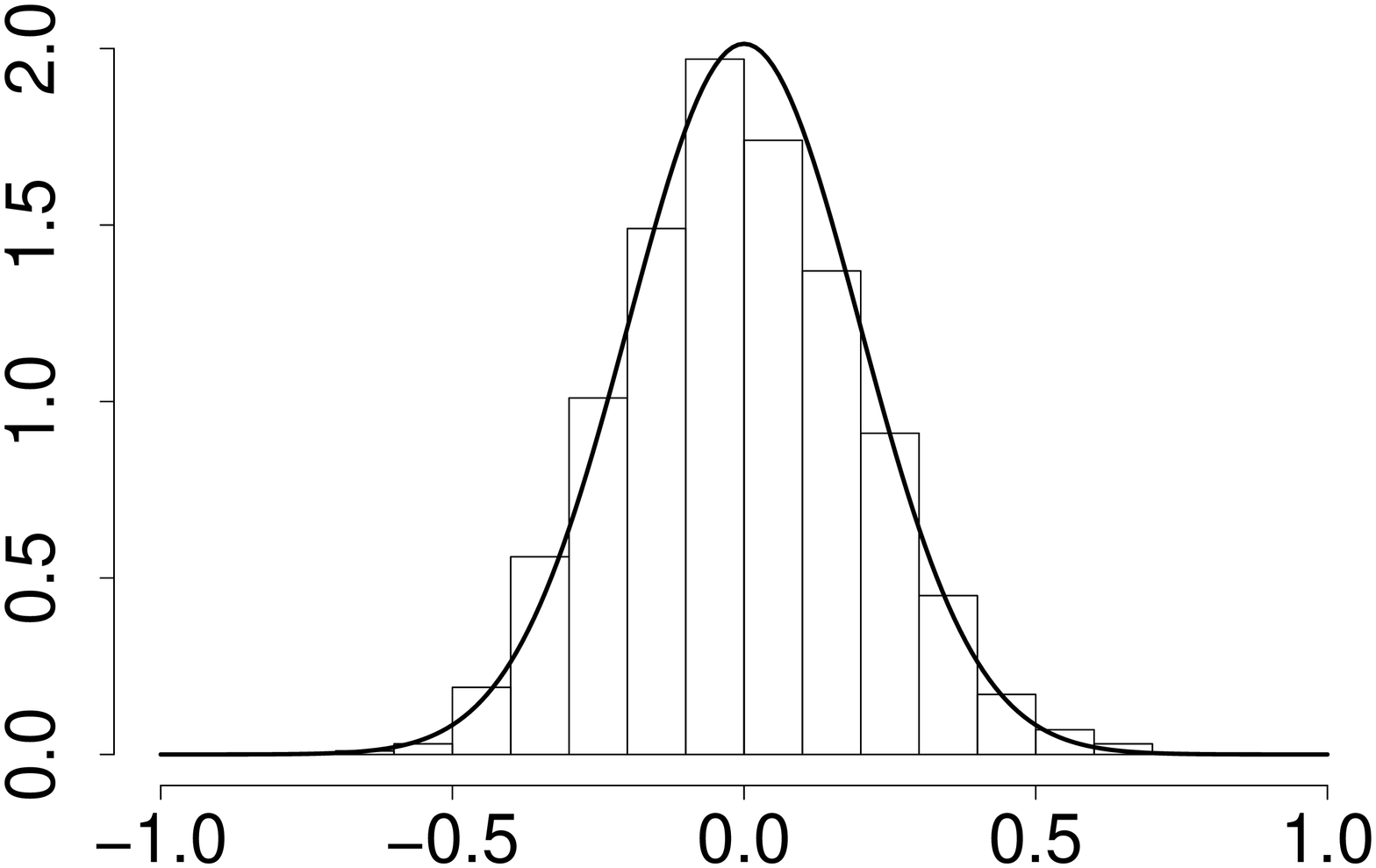}}&
\rotatebox{270}{\includegraphics[width=0.28\linewidth,height=0.22\linewidth]{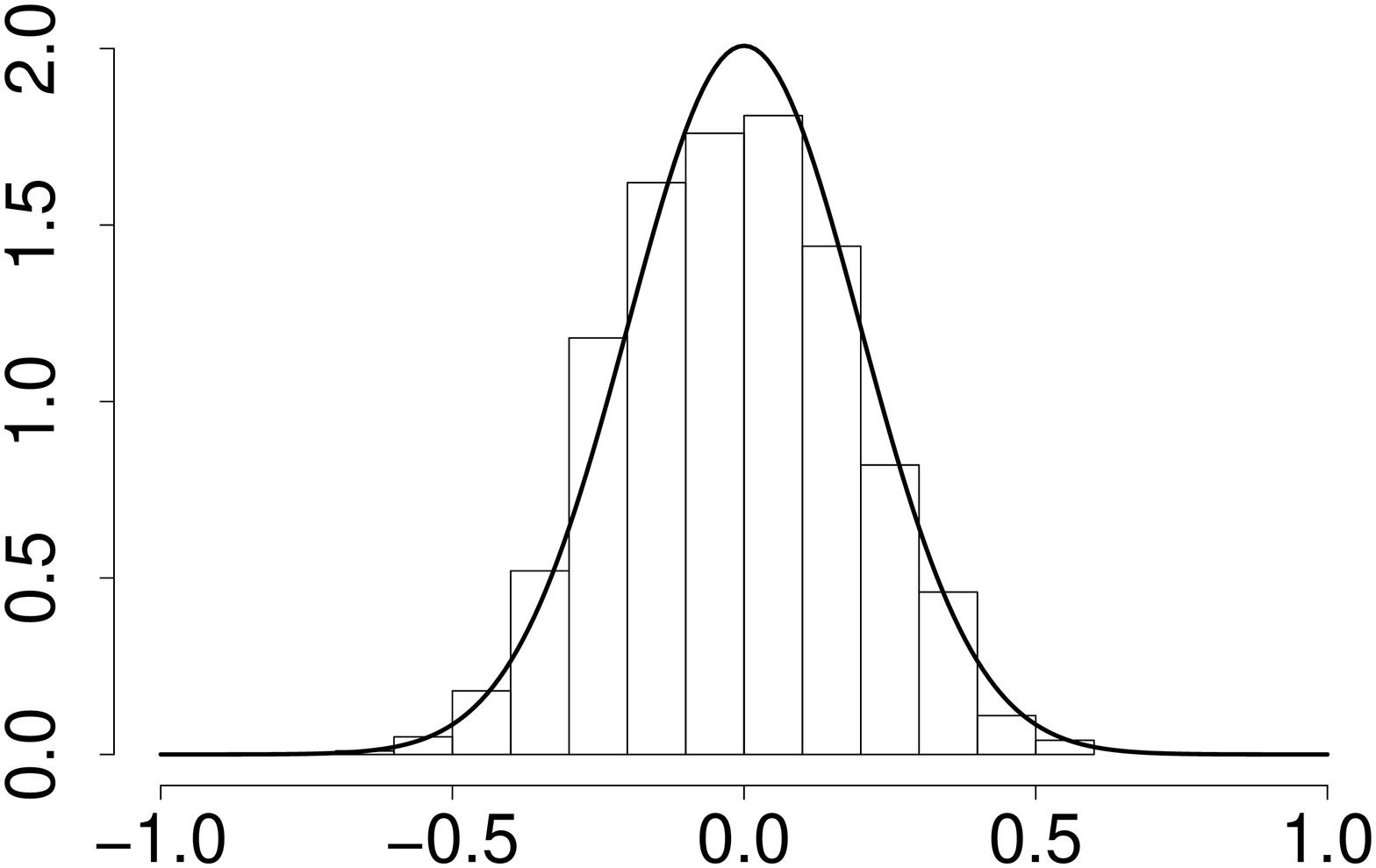}}
\end{tabular}
\end{center}
\caption{\sl Normal QQ-plots and histograms of the samples of $\CC_{ij|v,n}(0$$\cdot$$4,0$$\cdot$$2)$ at levels 3 and 4 of the Gaussian canonical and regular vines (column 1 and 2), respectively, with $\rho=0$$\cdot$$5$ and $n=1,000$, along with the corresponding conjectured limiting probability density functions}
\label{fig:simvar_qq_hist_CR_Gauss}
\end{figure}

\begin{figure}%[H]
\begin{center}
\begin{tabular}{c@{\quad}c}
\rotatebox{270}{\includegraphics[width=0.28\linewidth,height=0.22\linewidth]{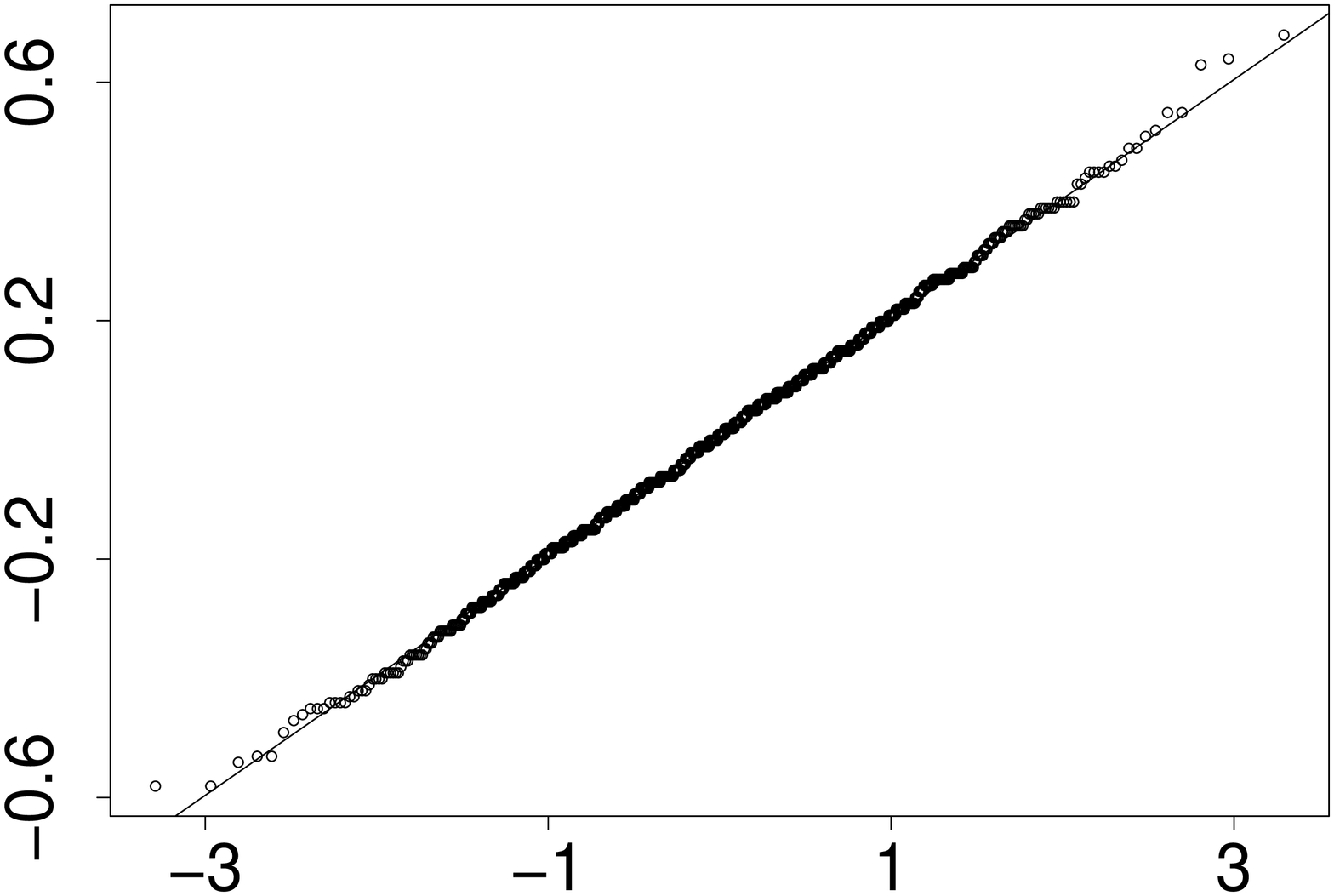}}&
\rotatebox{270}{\includegraphics[width=0.28\linewidth,height=0.22\linewidth]{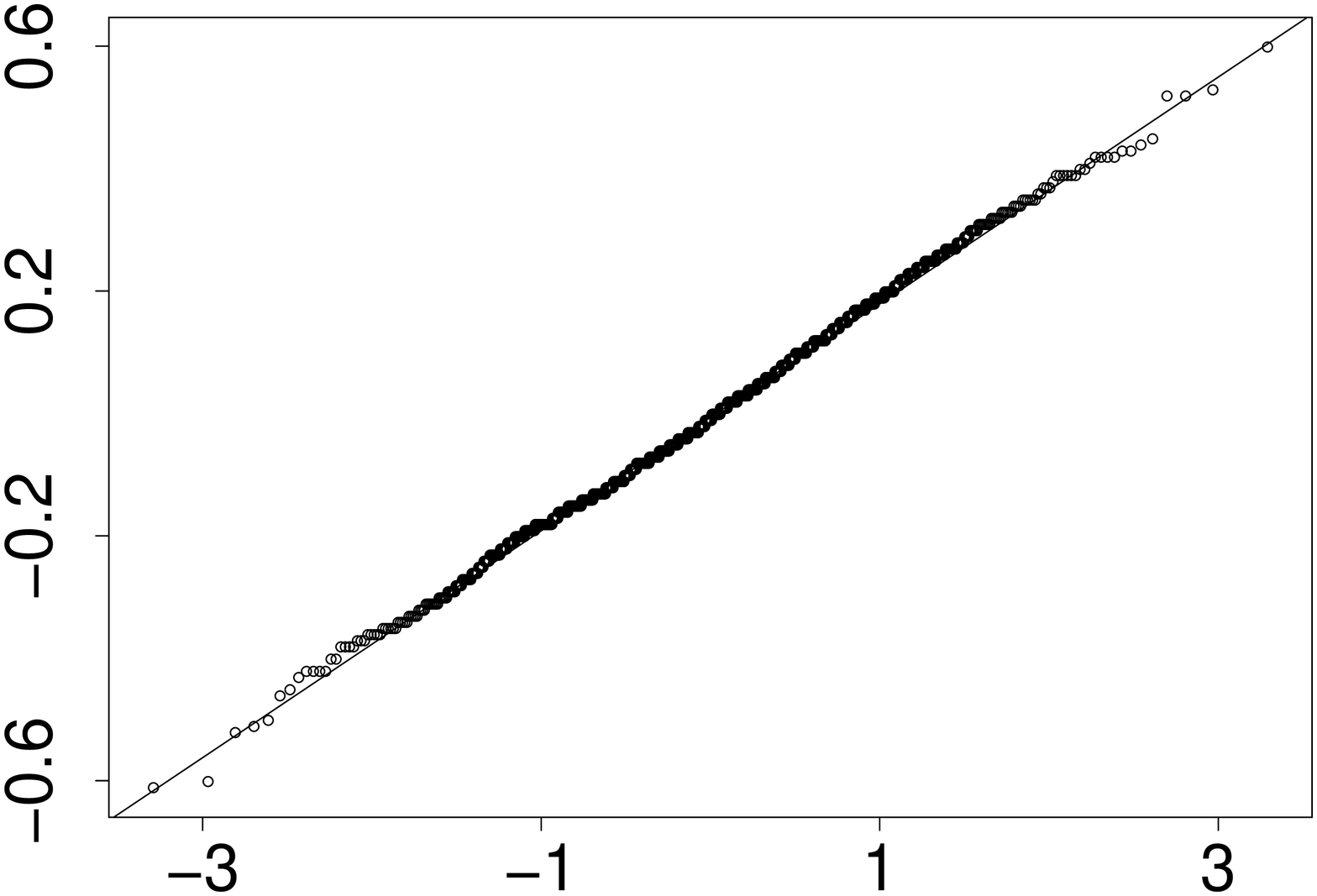}}\\
\rotatebox{270}{\includegraphics[width=0.28\linewidth,height=0.22\linewidth]{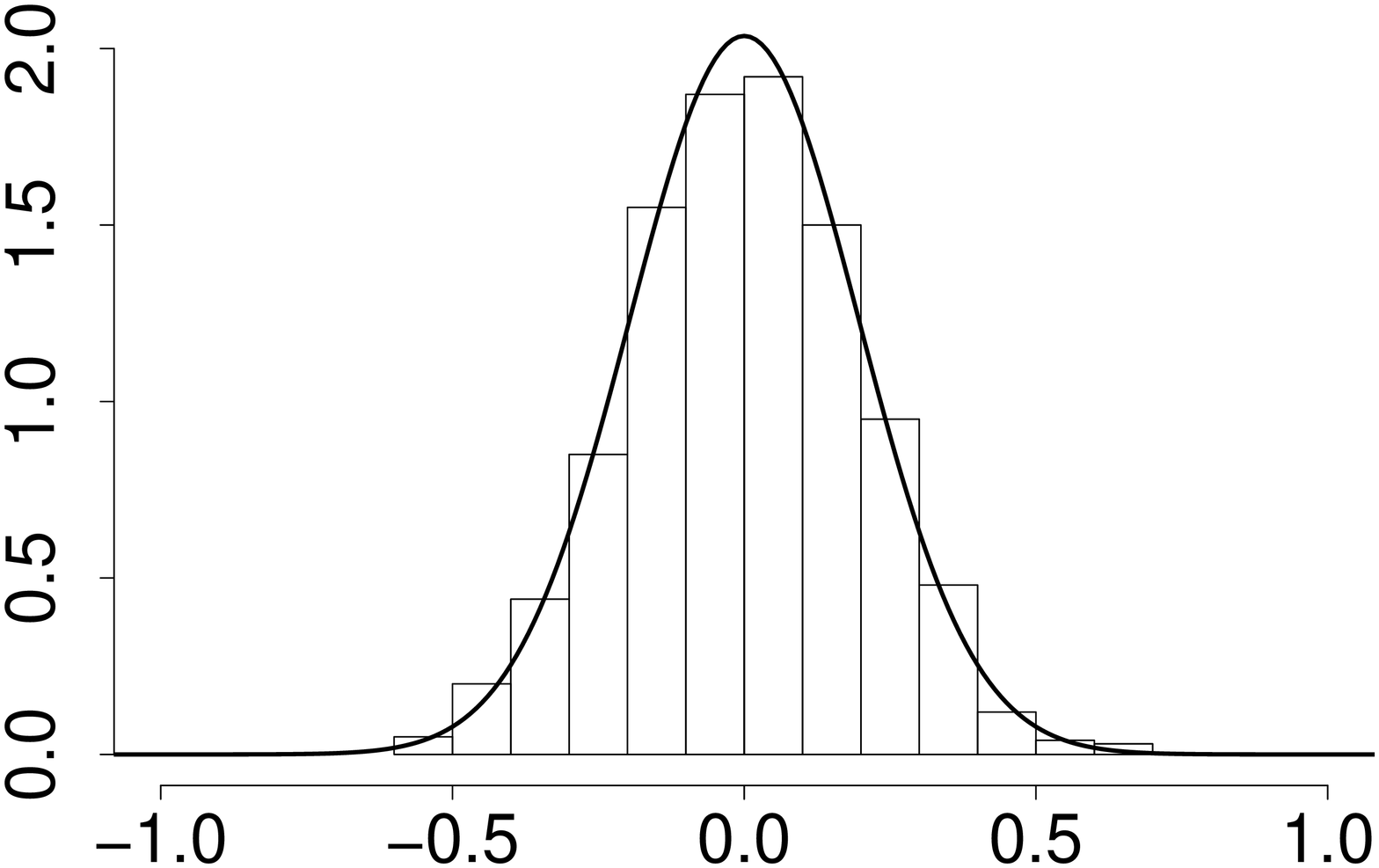}}&
\rotatebox{270}{\includegraphics[width=0.28\linewidth,height=0.22\linewidth]{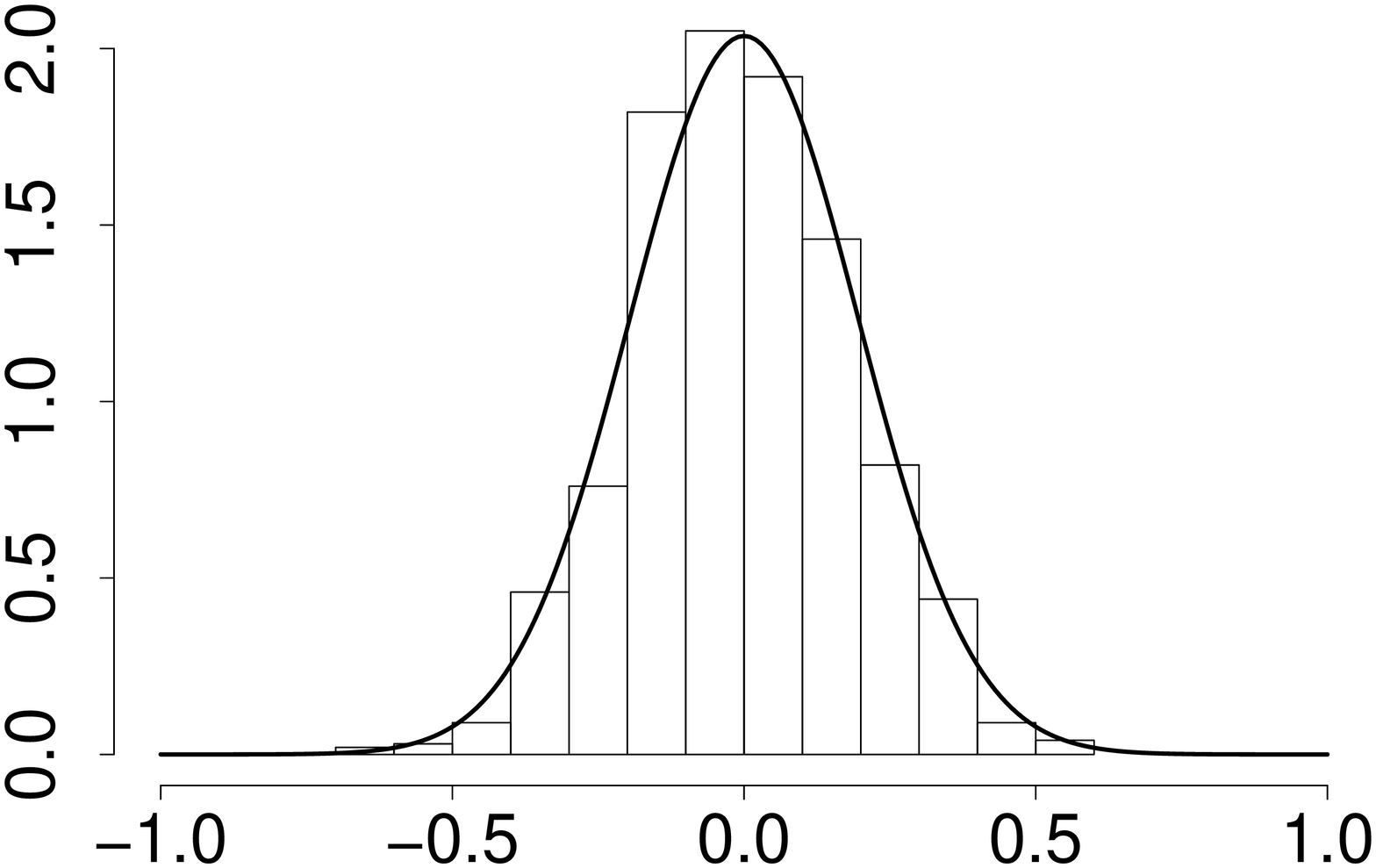}}
\end{tabular}
\end{center}
\caption{\sl Normal QQ-plots and histograms of the samples of $\CC_{13|2,n}(0$$\cdot$$4,0$$\cdot$$2)$ from the Student's t and a Gumbel drawable vines (column 1 and 2) with $n=1,000$ and $(\rho=0$$\cdot$$5,\nu=6)$ and $\alpha=1$$\cdot$$5$, respectively, along with the corresponding conjectured limiting probability density functions}
\label{fig:simvar_qq_hist_D_Student_Gumbel}
\end{figure}

\begin{figure}%[H]
\begin{center}
\begin{tabular}{c@{\quad}c@{\quad}c}
\rotatebox{270}{\includegraphics[width=0.28\linewidth,height=0.22\linewidth]{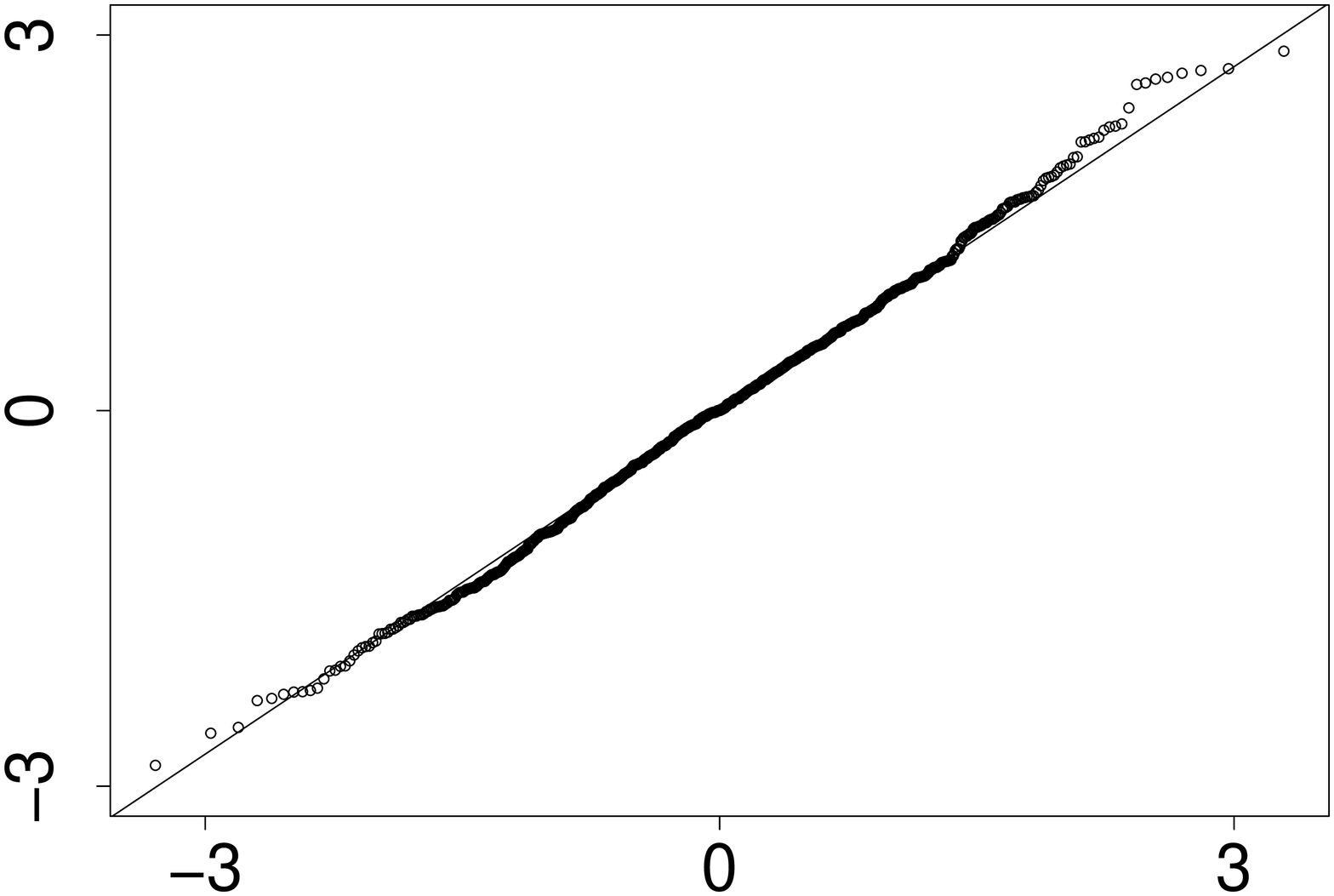}}&
\rotatebox{270}{\includegraphics[width=0.28\linewidth,height=0.22\linewidth]{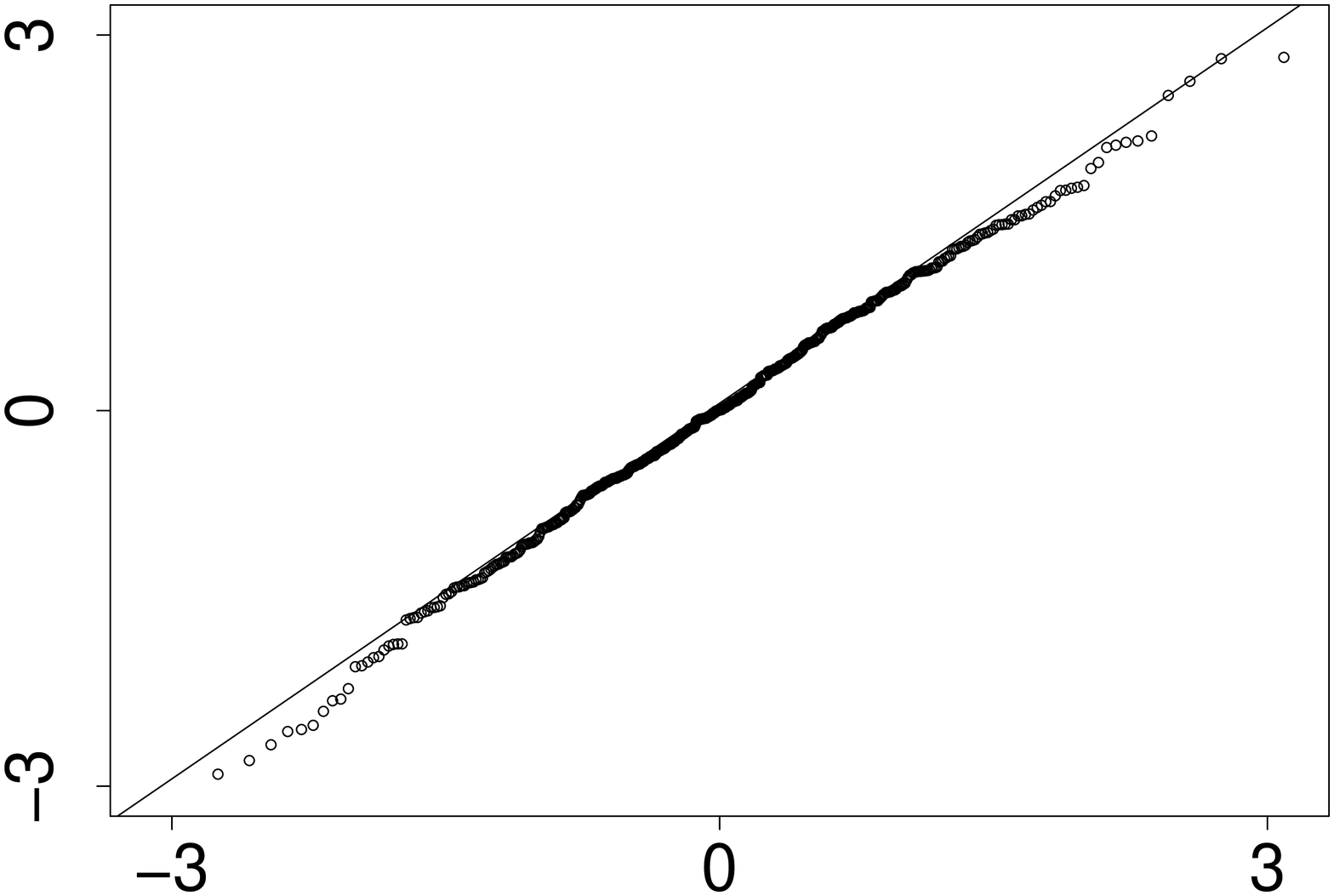}}&
\rotatebox{270}{\includegraphics[width=0.28\linewidth,height=0.22\linewidth]{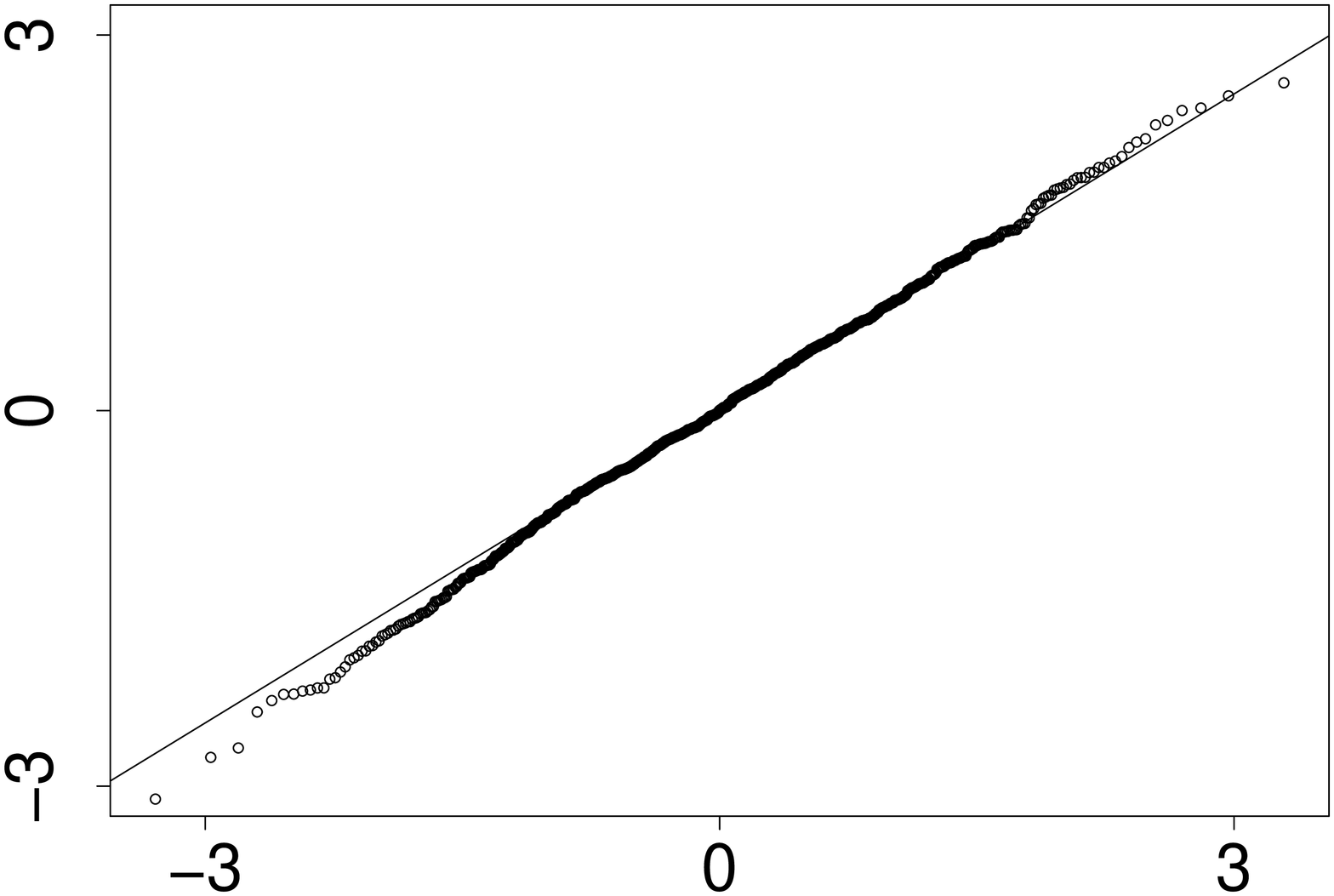}}\\
\rotatebox{270}{\includegraphics[width=0.28\linewidth,height=0.22\linewidth]{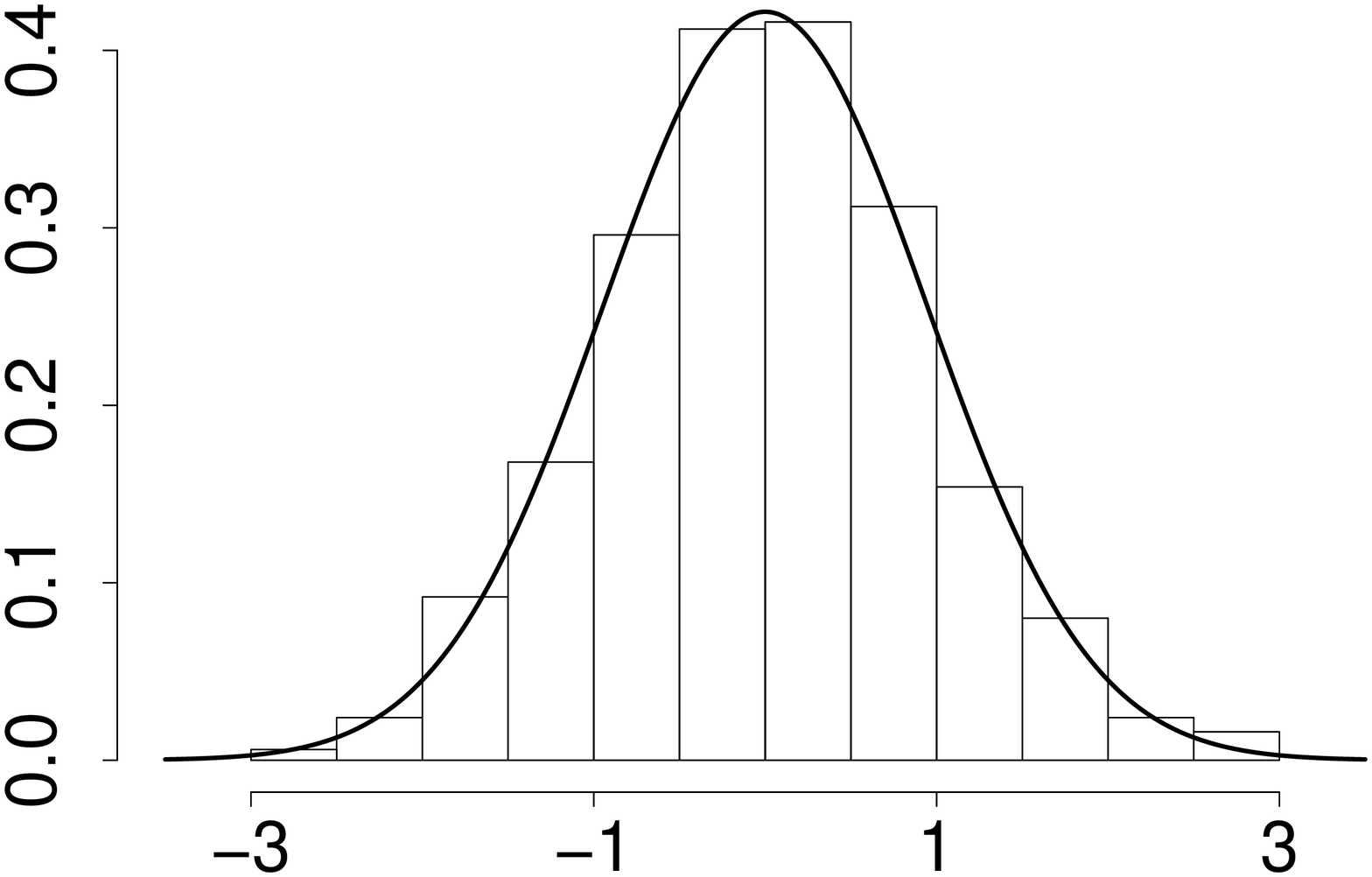}}&
\rotatebox{270}{\includegraphics[width=0.28\linewidth,height=0.22\linewidth]{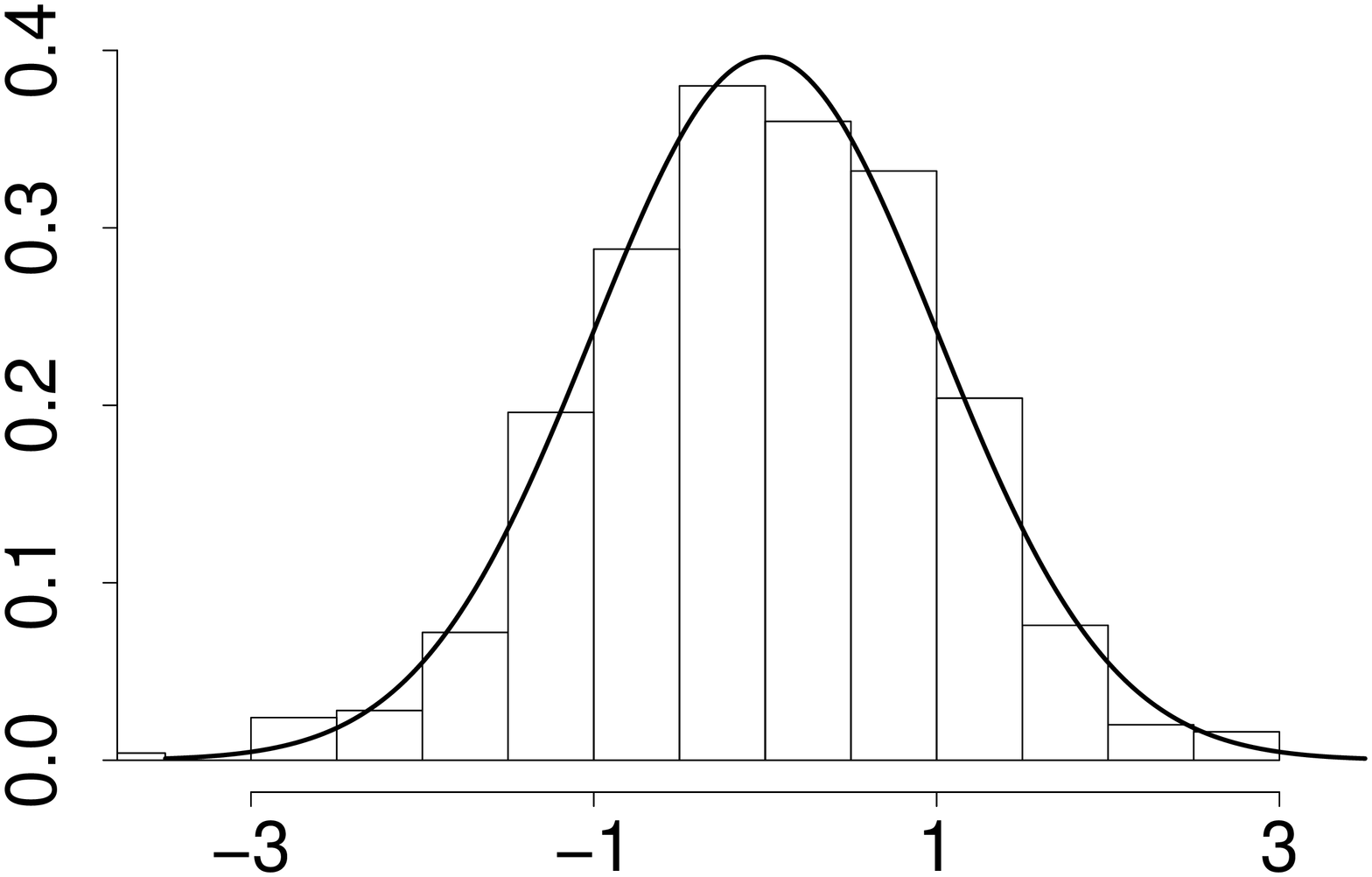}}&
\rotatebox{270}{\includegraphics[width=0.28\linewidth,height=0.22\linewidth]{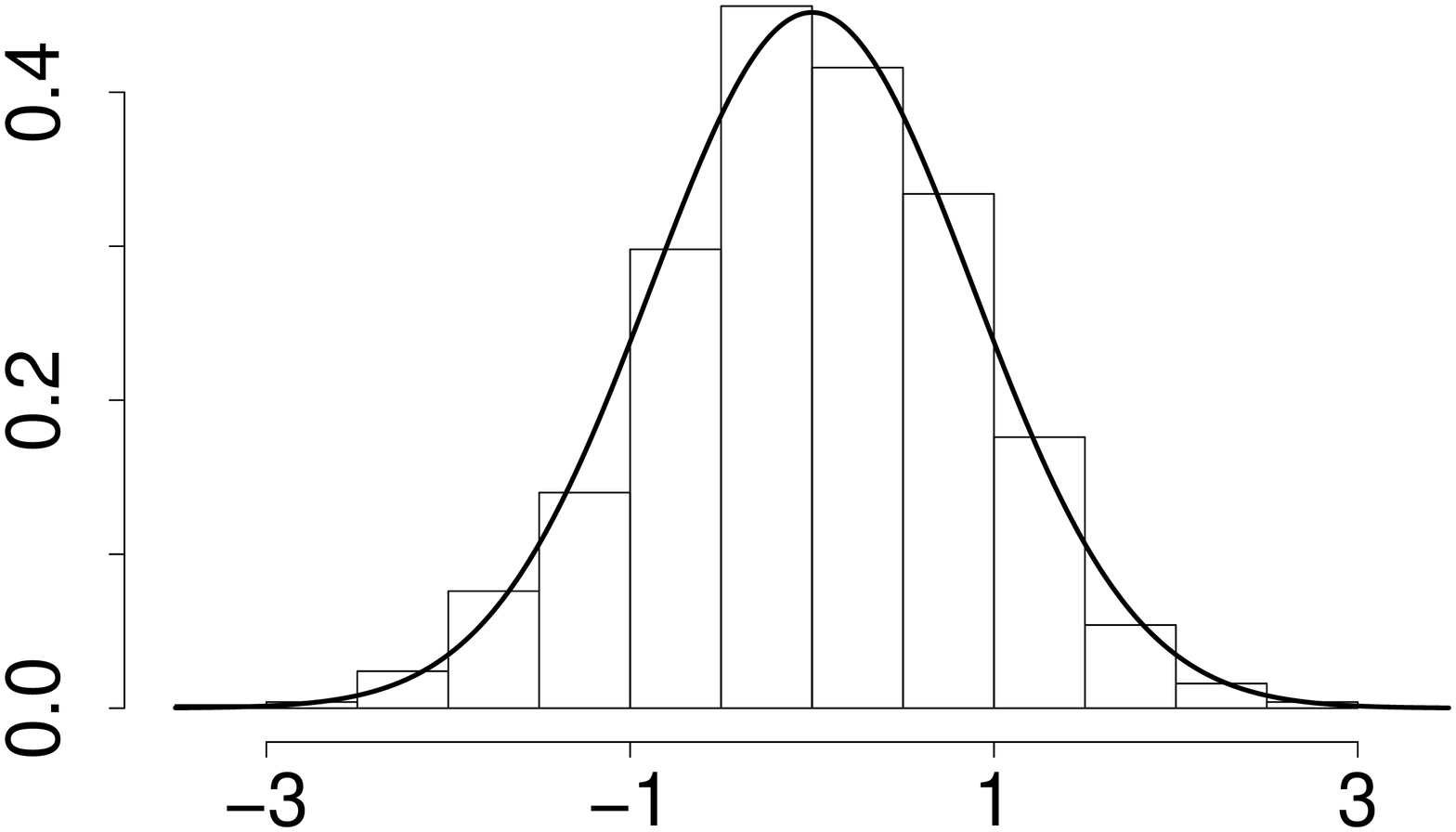}}
\end{tabular}
\end{center}
\caption{\sl Normal QQ-plots and histograms of the samples of $12 \int \CC_{14|23}$ from the Gaussian, Student's t and a Gumbel models with $n=1,000$ and $\rho=0$$\cdot$$5$, $(\rho=0$$\cdot$$5,\nu=6)$ and $\alpha=1$$\cdot$$5$, respectively, along with the corresponding conjectured limiting probability density functions}
\label{fig:simvar_qq_hist_rhoS}
\end{figure}

\begin{table}%[H]
\centering
\def~{\hphantom{0}}
\begin{tabular}{@{}lllllll}
\textsc{Model} & \textsc{$\alpha$} & \multicolumn{3}{c}{\textsc{Non-parametric}} & \multicolumn{2}{c}{\textsc{Parametric}}\\
 & & Percentile & Symmetric & Plug-in & Correct & Incorrect\\[5pt]
\multirow{3}{\LLLL}{Gaussian} & 0$\cdot$1 & 0$\cdot$91~(0$\cdot$99) & 0$\cdot$91~(0$\cdot$99) & 0$\cdot$92~(1$\cdot$0) & 0$\cdot$90~(0$\cdot$94) & 0$\cdot$86~(0$\cdot$93)\\
 & 0$\cdot$05 & 0$\cdot$95~(1$\cdot$2) & 0$\cdot$95~(1$\cdot$2) & 0$\cdot$95~(1$\cdot$2) & 0$\cdot$95~(1$\cdot$1) & 0$\cdot$92~(1$\cdot$1)\\
 & 0$\cdot$01 & 0$\cdot$99~(1$\cdot$5) & 0$\cdot$99~(1$\cdot$6) & 0$\cdot$99~(1$\cdot$6) & 0$\cdot$99~(1$\cdot$5) & 0$\cdot$97~(1$\cdot$5)\\
 \\
\multirow{3}{\LLLL}{Student's t} & 0$\cdot$1 & 0$\cdot$90~(1$\cdot$0) & 0$\cdot$90~(1$\cdot$0) & 0$\cdot$91~(1$\cdot$1) & 0$\cdot$89~(1$\cdot$0) & 0$\cdot$00~(0$\cdot$99)\\
 & 0$\cdot$05 & 0$\cdot$95~(1$\cdot$2) & 0$\cdot$95~(1$\cdot$2) & 0$\cdot$95~(1$\cdot$3) & 0$\cdot$93~(1$\cdot$2) & 0$\cdot$00~(1$\cdot$2)\\ 
 & 0$\cdot$01 & 0$\cdot$99~(1$\cdot$6) & 0$\cdot$99~(1$\cdot$6) & 0$\cdot$99~(1$\cdot$7) & 0$\cdot$97~(1$\cdot$6) & 0$\cdot$00~(1$\cdot$5)\\ 
\\
\multirow{3}{\LLLL}{Gumbel} & 0$\cdot$1 & 0$\cdot$92~(0$\cdot$88) & 0$\cdot$92~(0$\cdot$88) & 0$\cdot$92~(0$\cdot$97) & 0$\cdot$91~(0$\cdot$83) & 0$\cdot$68~(0$\cdot$80)\\
 & 0$\cdot$05 & 0$\cdot$96~(1$\cdot$0) & 0$\cdot$96~(1$\cdot$1) & 0$\cdot$96~(1$\cdot$2) & 0$\cdot$95~(0$\cdot$98) & 0$\cdot$79~(0$\cdot$95)\\
 & 0$\cdot$01 & 0$\cdot$99~(1$\cdot$4) & 0$\cdot$99~(1$\cdot$4) & 0$\cdot$99~(1$\cdot$5) & 0$\cdot$99~(1$\cdot$3) & 0$\cdot$91~(1$\cdot$2)\\
\end{tabular}
\caption{\sl Coverage and length (upper and lower value, respectively) of the estimated confidence intervals for $\rho(C_{14|23})$ in the Gaussian, Student's t and Gumbel models with $n=1,000$. The lengths are multiplied by $10$.}
\label{tab:resampling_rhoS}
\end{table}

\begin{figure}[H]
\begin{center}
\includegraphics[width=0.8\linewidth]{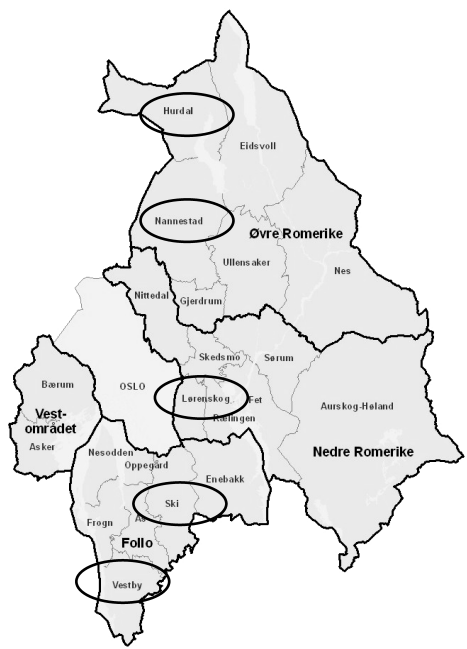}
\caption{\sl Meteorological stations where the precipitations in Section 5$\cdot$2 were recorded}
\label{fig:regnKart}
\end{center}
\end{figure}

\end{document}